\documentclass[10pt,twocolumn,twoside,submit]{JCNtran}
\pdfoutput=1
\usepackage{color}
\usepackage{amsmath}
\usepackage{amssymb}
\usepackage{tabularx}
\usepackage{epsfig}
\usepackage{graphicx}
\usepackage[T1]{fontenc}
\usepackage{enumitem}
\usepackage{multirow}
\usepackage{mathtools}

\usepackage[utf8]{inputenc}
\usepackage{textcomp}

\def\BibTeX{{\rm B\kern-.05em{\sc i\kern-.025em b}\kern-.08em
    T\kern-.1667em\lower.7ex\hbox{E}\kern-.125emX}}

\normalsize

\newtheorem{prop}{Proposition}
\newtheorem{theorem}{Theorem}

\newcommand{\myhl}[1]{#1} 

\newcommand{\V}[1]{\textup{#1}}

\begin{document}
\bibliographystyle{jcn}

\title{
Analysis of LTE-A Heterogeneous Networks with SIR-based Cell Association and Stochastic Geometry}
\author{Giovanni Giambene, Van Anh Le
\thanks{Manuscript received May 15, 2017 approved for publication by Young-Chai Ko, JCN Editor, November 28, 2017.}
\thanks{This research was supported by CNIT and the University of Siena.}
\thanks{The authors are with the Department of Information Engineering and Mathematical Sciences, University of Siena, Via Roma 56, 53100 Siena,  Italy. Email: giambene@unisi.it, anh.le@unisi.it.}
\thanks{G. Giambene is  the corresponding author.}
\thanks{Digital Object Identifier TBD}}
\markboth{ACCEPTED TO BE PUBLISHED ON THE JOURNAL OF COMMUNICATIONS AND NETWORKS}
{Giambene and Le: Analysis of LTE-A Heterogeneous Networks} \maketitle

\begin{abstract}
This paper provides an analytical framework to characterize the performance of \emph{Heterogeneous Networks} (HetNets), where the positions of base stations and users are modeled by spatial Poisson Point Processes (stochastic geometry). We have been able to formally derive outage probability, rate coverage probability, and mean user bit-rate when a frequency reuse of $K$ and a novel prioritized SIR-based cell association scheme are applied. A simulation approach has been adopted in order to validate our analytical model; theoretical results are in good agreement with simulation ones. The results obtained highlight that the adopted cell association technique allows very low outage probability and the fulfillment of certain bit-rate requirements by means of adequate selection of reuse factor and micro cell density. This analytical model can be adopted by network operators to gain insights on cell planning. Finally, the performance of our SIR-based cell association scheme has been validated through comparisons with other schemes in literature.
\end{abstract}


\vspace{10pt}
\section{\uppercase{Introduction}}
\label{sec:Intro}
\PARstart{M}{obile} network operators are experiencing a significant traffic demand increase because of the emergence of bandwidth-consuming applications, such as video streaming over new-generation mobile devices \cite{CiscoWhitePaper}. \emph{Long Term Evolution - Advanced} (LTE-A) \cite{giambene2014resource} systems have addressed this issue using both high-power nodes (macro cells) and low-power ones (small cells). These \emph{Heterogeneous Cellular Networks} (HetNets) \cite{chandrasekhar2008femtocell}, also denoted as multi-tier cellular systems, need \emph{Inter-Cell Interference Coordination} (ICIC) mechanisms in order to deal with co-tier and cross-tier interference \cite{RefEMTC},\cite{icc}. The disparity between macro and micro cell transmission powers causes a load imbalance \cite{loadbalancingTrans}. Hence, it is important to offload traffic from macro to micro cells in order to improve user experience \cite{overviewloadbalancing}. In addition, the emergence of \emph{Fifth-Generation} (5G) cellular networks will lead to ultra-dense cell deployments to meet the new capacity needs. Thus, interference and load imbalance issues are becoming more critical to network performance.

The commonly-used hexagonal arrangement of base stations has some limitations in modeling cellular networks. In the reality, the positions of base stations do not exactly follow a hexagonal regular layout because of both variable traffic demands among different locations (e.g., rural, urban) and obstacles (e.g., mountains, forests, etc.) \cite{TractableApproach}. Since it is difficult to achieve an analytical model in realistic conditions \cite{WynerAccuracy},\cite{StochasticSurvey}, simulations are basically the only approach for performance evaluation in terms of capacity and outage probability.

On the other hand, stochastic geometry is a very powerful mathematical tool for modeling wireless networks with random topologies \cite{StochasticGeoandApplications},\cite{RandomPlaneNetwork}. The most famous point process applied to the study of HetNets is the \emph{Poisson Point Process} (PPP) \cite{StochasticGeoandApplications}. The PPP model was first proposed in \cite{TractableApproach} to describe the distribution of cells in classical cellular networks. Then, this study was extended to HetNets in \cite{HetNEtFlexibleAssoc}. The significant advantage in using PPP processes is that they can help to obtain closed-form formulas of important performance metrics, such as outage, average capacity, etc.

\subsection{Related Works}
\label{sec:relatedWorks}

In the literature, the papers using stochastic geometry analysis usually focus on two alternative cases, which are \emph{spectrum partitioning} \cite{AUtilityPerspective}-\cite{RateCoverate} and \emph{spectrum sharing} schemes \cite{HetNEtFlexibleAssoc},\cite{OffloadinginHetNets},\cite{recentpaper}. With spectrum sharing, the entire spectrum is reused in every cell; instead, the bandwidth is divided into different parts to be reused among cells in spectrum partitioning. For example, in two-tier HetNets, the bandwidth is divided into $F_1$ and $F_2$ segments with spectrum partitioning. There are two ways to use $F_1$ and $F_2$ as follows: (\emph{i}) The macro cell tier uses $F_1$, while the micro cell tier uses $F_2$; (\emph{ii}) The macro cell tier uses only $F_1$, while the micro cell tier can use both $F_1$ and $F_2$: $F_2$ is used by micro cells for biased \emph{User Equipments} (UEs) only (i.e., those UEs, originally belonging to the macro cell tier, which are forced to associate with the micro cell tier) according to a traffic offloading scheme \cite{rangeExpansion},\cite{Qualcomm}. Spectrum sharing and spectrum partitioning schemes cause UEs to suffer from co-tier and cross-tier interference.

Most of the papers adopting a stochastic geometry model assume a cell association scheme based on maximum received power (or maximum-biased received power) \cite{HetNEtFlexibleAssoc}-\cite{recentpaper}. However, forcing UEs to associate with the cell providing the maximum (or the maximum-biased) received power does not necessarily mean having \emph{Signal-to-Interference plus Noise Ratio} (SINR) higher than a minimum SINR threshold so that UEs may experience outage. The works in \cite{AnalysisMaxSIR}-\cite{AnalysisMaxSINRConnectivity} provide an analytical framework to study the performance of HetNets with a max-SINR cell association scheme. However, these papers adopt the simplified assumption of frequency reuse of 1. Thus, co-tier and cross-tier interference cause a large number of UEs to experience outage conditions. Moreover, none of these papers takes the load balance issue into account so that the macro tier has associated more UEs than the micro tier. The survey paper \cite{ultimo} deals with stochastic geometry and frequency reuse, but in a single-tier scenario, thus oversimplifying the model without cross-tier interference and load balancing issues.
Finally, the paper \cite{dopoultimo} uses reduced power transmissions for macrocells in some sub-frames for a two-tier system with PPP model. The association of UEs to micro-cells is privileged by means of a range expansion approach. The limit of this work is that only full frequency reuse is considered.

\subsection{Contributions and Organization}

In this study, we refer to downlink since it is more critical than uplink in terms of traffic demand. We consider a scenario with \emph{Frequency Reuse} (FR) of $K$, so that the entire spectrum is divided into $K$ equal-size parts. In particular, each macro/micro cell in the system randomly uses one frequency band to reduce the interference from other cells. Moreover, we consider a cell association scheme based on the SINR at the UEs: we propose a SINR-based cell association scheme, where the micro tier has higher priority in the association to achieve load balancing with the macro tier. In particular, a UE associates with a micro cell as long as it experiences SINR greater than a minimum SINR threshold $T$ from the micro tier; if the UE is in the outage area of the micro cell tier, it will consider to associate with a macro cell.

Even if a SINR-based cell association is a common approach in cellular systems, its analysis in the PPP stochastic geometry case is not so common in the literature. We address this issue and we provide analytical derivations on outage probability, average cell load, and rate coverage probability, representing the probability that the UE bit-rate is bigger than a certain threshold $R_T$. On the basis of input parameters such as bit-rate threshold $R_T$, minimum SINR threshold $T$, transmission powers of base stations, and UEs density, the purpose of this study is to determine the $K$ value and the base station densities that allow to fulfill requirements in terms of outage probability (see Section III) and rate coverage probability (see Section IV). A simulation approach has also been provided in order to validate our analysis.

The original contributions of this work can be summarized as follows:
\begin{itemize}[leftmargin=.2in]
\item The work in \cite{AnalysisMaxSIR} adopts a max-SINR cell association scheme without differentiation between macro and micro cells. This approach tends to underutilize the resources of the micro cell layer. In our work, we remove this issue by giving a strict priority to the micro cells for the UE cell selection process (cell offloading). Only when there is no micro cell available to serve the UE, it is associated with a macro cell.
\vspace{0.1cm}\item The works in \cite{HetNEtFlexibleAssoc},\cite{OffloadinginHetNets},\cite{recentpaper},\cite{AnalysisMaxSIR} adopt a full frequency reuse system. We remove such limitation so that our analysis considers $K$ frequency segments that can be assigned at random to both macro and micro cells.
\vspace{0.1cm}\item On the basis of the analysis provided in this paper, a cell planning optimization approach is proposed to select the reuse factor $K$ and the ratio of micro-to-macro cell densities, depending on the ratio of micro-to-macro transmission power levels and other system parameters.\\
\end{itemize}

The rest of this paper is organized as follows: Section \ref{sec:ModelandAssumptions} presents system model and assumptions for the study of two-tier HetNets with prioritized SINR-based cell association. In Section \ref{sec:OutageAverage}, we derive the outage probability and average load on each tier. Section \ref{sec:Rate} provides the analysis of rate coverage probability and mean UE bit-rate. Section \ref{sec:Simulation} shows the simulation approach adopted in this study along with the settings of the LTE-A-based HetNet system. Results are presented in Section \ref{sec:ResultsPPP}, followed by Section \ref{sec:Conclusions} that provides the conclusions.

\vspace{10pt}
\section{\uppercase{System Model and Assumptions}}
\label{sec:ModelandAssumptions}

\subsection{Scenario Description}

\begin{figure}
 \centering \epsfxsize=9cm \epsfbox{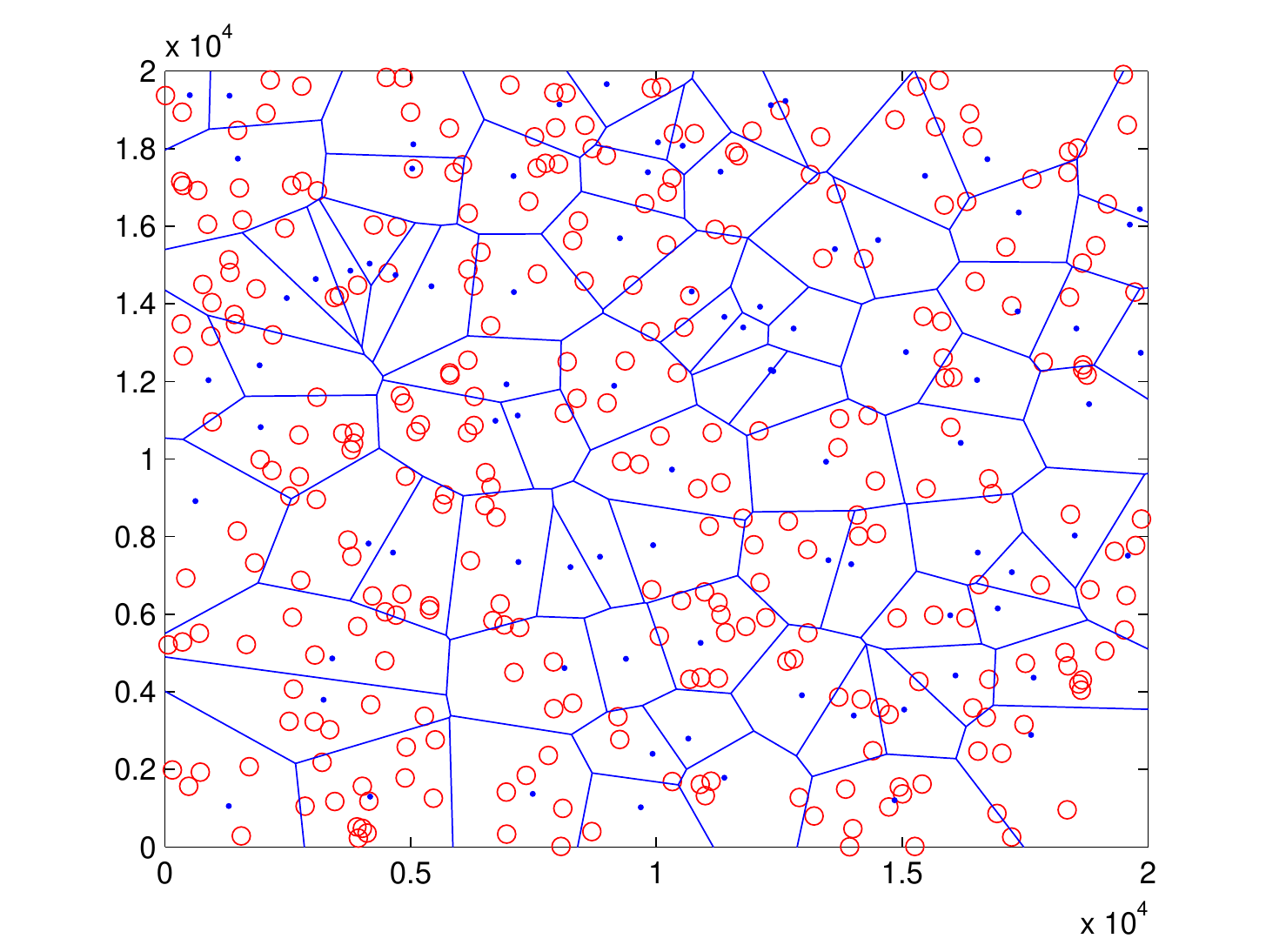}
 \caption{Heterogeneous network with PPP distribution.}\label{fig:PPPScenario}
\end{figure}

Let us consider a two-tier HetNet scenario, where macro cells, micro cells, and UEs are placed in the service area according to independent homogeneous PPPs. In particular, let $\Phi_{M}$ with density $\lambda_{M}$, $\Phi_{\mu}$ with density $\lambda_{\mu}$, and $\Phi_{u}$ with density $\lambda_{u}$ denote the PPPs characterizing macro eNBs (M-eNBs), micro eNBs (\textmu-eNBs), and UEs, respectively; the densities represent the average number of points of the processes per area unit. $\gamma_{M}$ and $\gamma_{\mu}$ denote the path loss exponents for macro and micro cell layers. For the sake of simplicity, we assume $\gamma_{M} = \gamma_{\mu} = \gamma$. Moreover, M-eNBs use the  transmission power $P_{M}$ that is higher than the transmission power $P_{\mu}$ used by \textmu-eNBs. The system bandwidth is denoted by $W$. Let $h$ denote the channel gain (power factor) due to Rayleigh fading; $h$ has an exponential distribution with unitary mean. Vector $r_i$ denotes the location of eNB $i$ (being it a M-eNB or a \textmu-eNB) assuming that a reference UE is in the origin. Fig. \ref{fig:PPPScenario} shows an example of HetNet scenario according to our PPP assumption: dots indicate M-eNBs, while circles are \textmu-eNBs. The lines among macro cells are obtained by using a Voronoi diagram for the macro tier, however, they do not reflect the actual cell association scheme adopted in our study (i.e., these lines are not real cell borders in this case).

\begin{figure}
  \centering \centering \epsfxsize=6.5cm \epsfbox{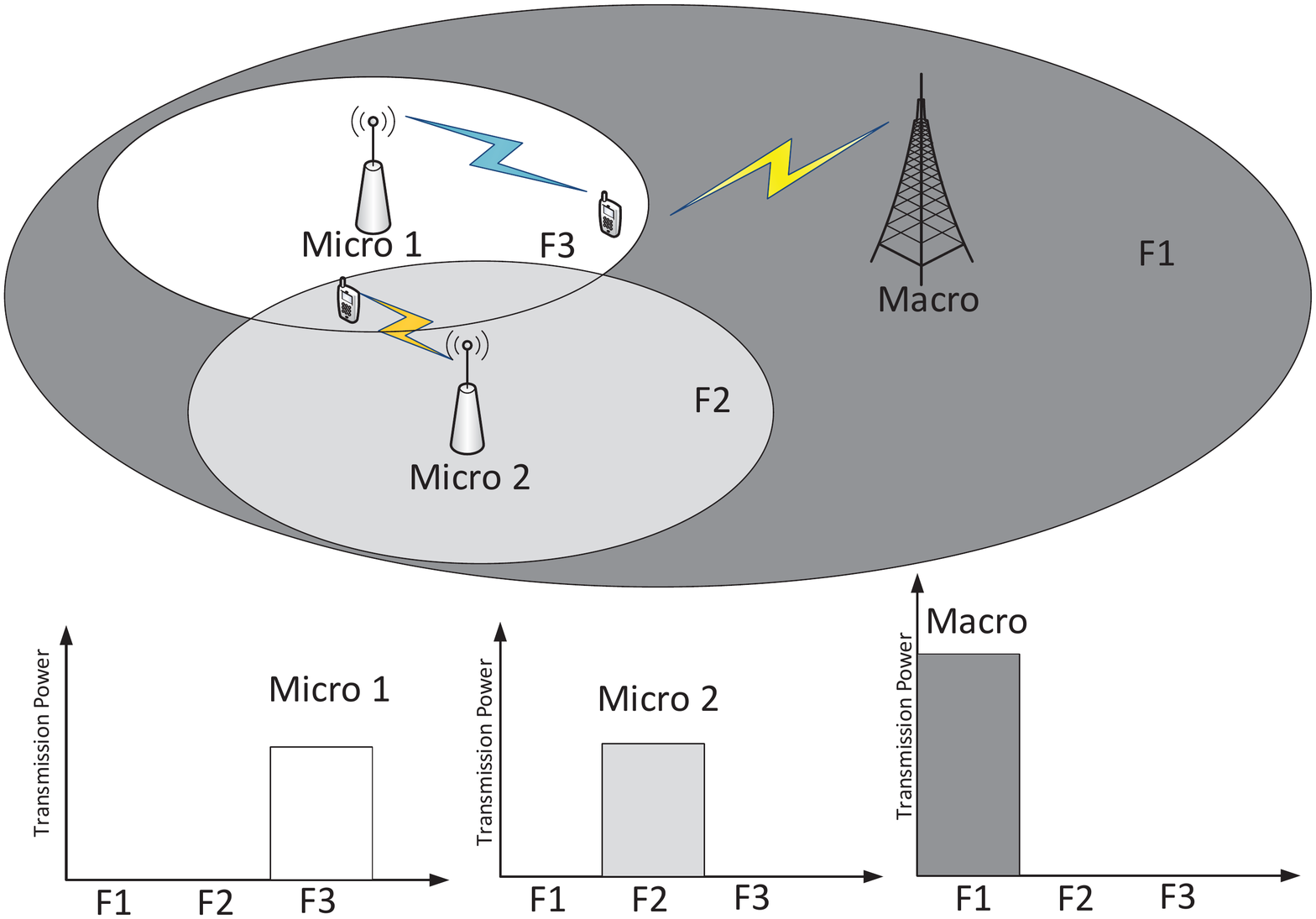}
  \caption{FR scheme adopted in the HetNet scenario for $K$ = 3.}\label{fig:FFRHetNet}
\end{figure}

In this study, because of the dense deployment of eNBs, we neglect the background noise with respect to the interference to simplify the analysis \cite{AUtilityPerspective}; therefore, from now on SINR will simply become SIR. In this scenario, FR of $K$ is adopted to improve SIR of UEs, especially for those in edge areas, and to reduce outage probability. The entire frequency band is divided into $K$ equal segments, denoted as $\{F_1, F_2,...,F_K\}$. Moreover, we assume that each cell randomly selects 1 out of the $K$ segments. An example of FR with $K = 3$ is shown in Fig. \ref{fig:FFRHetNet}, where the macro cell selects $F_1$ and the two micro cells use $F_2$ and $F_3$, respectively. As we can see, interference is significantly reduced in this case with respect to spectrum sharing schemes.

\subsection{SIR Model}
Assuming that the reference UE is located in the origin and associates with cell $i$ using frequency segment $k$, where $k \in \{1,2,...,K\}$, its SIR can be expressed as follows:

\begin{equation}\label{eq:SINRmacro}
  \V{SIR}_i = \frac{P_i h_i \parallel r_i \parallel^{-\gamma}}{\sum_{j \in \left\{\Phi_{M,k} \cup \Phi_{\mu,k}\right\} \setminus i} P_j h_j \parallel r_j \parallel^{-\gamma}},
\end{equation}
where $\Phi_{M,k} \cup \Phi_{\mu,k}$ denotes the sets of M-eNBs and \textmu-eNBs using frequency segment $k$, $\parallel r_i \parallel$ denotes the distance from eNB $i$ to the reference UE. $P_i$ can be either $P_M$ or $P_{\mu}$ depending on $i$ being a macro cell or a micro cell. Note that the background noise has been neglected in this formula as explained in previous sub-Section. Assuming that there is mapping from cell index $i$ and frequency segment $k$ assigned to cell $i$, index $k$ has been omitted from the above SIR notation.

\vspace{10pt}
\section{\uppercase{Outage Probability}}
\label{sec:OutageAverage}

\subsection{Outage Probability Analysis}
\label{subsec:OutageProb}
Let $T$ denote the minimum SIR threshold. A UE is not in outage conditions if it experiences SIR higher than $T$ in at least one cell of the system. The outage probability is given by $ {O} = 1 -  {P}_c$, where $ {P}_c$ denotes the coverage probability of the entire network. A UE is in the coverage area of an arbitrary cell $i$ (could it be a M-eNB or a \textmu-eNB) in the network if $\V{SIR}_i > T$. Thus, the coverage probability of the entire network including all macro and micro cells can be expressed as follows:

\begin{equation}\label{eq:coverageProb}
   {P}_c =  {P}\left(\bigcup\limits_{i \in \Phi_{M} \cup \Phi_{\mu}}\left\{\V{SIR}_i > T\right\}\right),
\end{equation}
where $\Phi_{M}$ and $\Phi_{\mu}$ indicate the sets of macro and micro cells: $i \in \Phi_M \cup \Phi_{\mu}$ denotes one point (and then one eNB) belonging to the sum process $\Phi_M \cup \Phi_{\mu}$. This probability is difficult to obtain since it involves calculating the coverage probability of every cell (i.e., $ {P}\{ \V{SIR}_i > T \}$) and then to consider the union. Note that a UE could be in the coverage areas of different cells at the same time, so that their union is not empty. In order to deal with this issue, we consider $ {P}_{c,k} =  {P} (C_k)$, the probability of event $C_k$ that the reference UE is in the coverage area of at least one of the cells that use frequency $F_k$ (considering all together micro and macro cells using $F_k$). In other words, we study the coverage event $C_k$ of each frequency segment, considering all the tiers together; then, we take the union of the coverage events of all frequency segments. Thus, we have:

\begin{equation}\label{eq:coverageProb2}
   {P}_c =  {P} \left( \bigcup\limits_{k \in \{1,2,...,K\}} C_k \right).
\end{equation}
By using the inclusion-exclusion property, we obtain:
\begin{equation}\label{eq:in-exclusion}
\begin{split}
    &   {P} \left( \bigcup\limits_{k \in \{1,2,...,K\}} C_k \right) =\\
     &  \sum_{k=1}^{K}\left( (-1)^{k+1} \sum_{x_1<x_2<...<x_k}  {P}(C_{x_1}\cap C_{x_2}\cap ... \cap C_{x_k}) \right),
\end{split}
\end{equation}
where $x_i \in \{1,2,...,K \}$. According to (\ref{eq:in-exclusion}), for a given $k$, we select all possible sets of  $x_i$ values satisfying the condition ${x_1<x_2<...<x_k}$. For example, when $K = 3$, we have:
\begin{equation}\label{eq:coverageProb3}
\begin{split}
     {P}_c & =  {P} (C_1) +  {P} (C_2) +  {P} (C_3) \\
     & - \left[ {P}(C_1\cap C_2)+ {P}(C_2\cap C_3)+ {P}(C_1\cap C_3)\right] \\
     & +  {P}(C_1\cap C_2\cap C_3) \\
     & =  {P}_{c,1} +  {P}_{c,2} +  {P}_{c,3}  -  {P}(C_1\cap C_2) - {P}(C_2\cap C_3) \\
     & -  {P}(C_1\cap C_3) +  {P}(C_1\cap C_2\cap C_3).
\end{split}
\end{equation}
 The difficulty is to compute the terms $ {P}(C_1\cap C_2)$, $ {P}(C_2\cap C_3)$, $ {P}(C_1\cap C_3)$, and $ {P}(C_1\cap C_2\cap C_3)$. It is important to note that with FR of $K$, cells using frequency $F_x$ do not interfere with cells using frequency $F_y$ when $x \neq y$, thus $C_x$ and $C_y$ are totally independent events; nevertheless, coverages of $C_x$ and $C_y$ can have some degree of overlap (their intersections can be non-empty)\footnote{We will not consider overlap among cells using the same frequency. As commented in Appendix A, this is true for SINR threshold $T \ge 1$; instead, this is approximate for  $T < 1$.}. On the basis of this consideration, we can re-write the expressions in (\ref{eq:coverageProb3}) as follows:

\begin{equation}\label{eq:jointProb1}
   {P}(C_x\cap C_y) =  {P}(C_x)  {P}(C_y) =  {P}_{c,x} {P}_{c,y}
\end{equation}
and similarly
\begin{equation}\label{eq:jointProb1}
   {P}(C_1\cap C_2\cap C_3) =  {P}_{c,1} {P}_{c,2} {P}_{c,3}.
\end{equation}

Moreover, due to the symmetry of the problem and assuming to have a sufficiently-large number of cells, we have $ {P}_{c,1}= {P}_{c,2}= {P}_{c,3} = ...= {P}_{c,K}$, thus, formula (\ref{eq:in-exclusion}) can be further simplified by using the Newton Binomial Theorem as follows:
\begin{equation}\label{eq:coverageProb4}
    {P}_c  = \sum_{k=1}^{K}  (-1)^{k+1} \binom{K}{k}  {P}_{c,1}^k  = 1 - (1- {P}_{c,1})^K.
\end{equation}
\myhl{From Theorem \ref{theremOutage} below, we obtain the expression of $ {P}_{c,1}$ in the special case when both macro and micro tiers have the same SIR threshold $T$ and noise is neglected (a more general expression of $ {P}_{c,1}$ is provided in Appendix A):
\begin{equation}\label{eq:theorem1}
     {P}_{c,1} \approx D(\gamma, T) \triangleq \frac{\pi}{C(\gamma)T^{2/\gamma}},
  \end{equation}
}
where $C\left( \gamma  \right) = \frac{{2\pi ^2 }}{{\gamma \;\sin \left( {\frac{{2\pi }}{\gamma }} \right)}}$ and where probability $D(\gamma, T)$ is defined in (\ref{eq:theorem1}) itself.

In what follows, we will also use notations like $P_{c}(T)$ and $P_{c,1}(T)$ to stress that these quantities depend on SIR threshold $T$.

Finally, the outage probability can be expressed as follows by means of (\ref{eq:coverageProb4}):
\begin{equation}\label{eq:coverageProbFinal}
     {O} = 1 -  {P}_c = (1- {P}_{c,1})^K.
\end{equation}

\begin{theorem}
\label{theremOutage}
  \textit{The coverage probability $ {P}_{c,1}$ when only cells with frequency $F_1$ are considered, assuming that macro and micro tiers have SINR threshold $T_M$ and $T_{\mu}$ respectively, can be expressed as}
  \begin{equation}\label{th:generalizedCoverageProb}
  \begin{split}
     &  {P}_{c,1}(T_M, T_{\mu}) \approx \sum_{\underset{}{j = \{M, \mu\}}} \frac{\lambda_j}{K} \times \\
     &\int_{ {\mathbb{R}}^2} \exp\left[ -\left(\frac{T_j}{P_j}\right)^{2/\gamma} \parallel r_j \parallel^2 C(\gamma) \sum_{m = \{M,\mu\}} \frac{\lambda_m}{K} P_m^{2/\gamma}\right] \times \\
     & \exp\left(\frac{-T_j \sigma^2 \parallel r_j \parallel^\gamma}{P_j}\right) \V{d}r_j,
  \end{split}
\end{equation}
  where notations are detailed in Appendix A.
\end{theorem}

This Theorem is adapted from Theorem 1 in \cite{AnalysisMaxSIR} taking frequency reuse into account. The proof of this Theorem is provided in Appendix A. In this theorem, we only refer to (both macro and micro) cells using frequency $F_1$ out of all the cells using frequencies $F_1$, $F_2$, ..., $F_k$. In Appendix A, we also explain the approximation we made to achieve the result in (11).

\subsection{Average Cell Load}

The average cell load represents the fraction of active UEs belonging to the micro or the macro tier or, equivalently, the probability that a random UE belongs to a tier given that this UE is under the coverage area of the network (i.e., not in outage condition). Cells with a higher number of associated UEs will have a higher average load. Usually, UEs experience the best SIR from the macro tier, causing this tier to be overloaded if a common max-SIR cell association scheme is adopted. In order to achieve a better load balancing between micro and macro cells, we adopt a prioritized SIR-based cell association as follows: the micro tier has higher priority than the macro tier when cell association is performed. In particular, a UE associates with the micro tier as long as it experiences $\V{SIR}>T$ from this tier. Only if the UE has $\V{SIR}<T$ (outage) from the micro tier, it will consider to associate with the macro tier if it can guarantee $\V{SIR}>T$; otherwise, the UE experiences outage. This scheme allows reducing the number of UEs in macro cells while keeping the outage probability as low as possible.

The rationale of this scheme is that if a UE can be served by both macro and micro tiers, we prefer that this UE be associated with the micro tier every time this is possible, thus avoiding to use macro cell resources that could be more useful in those cases where UEs can only be covered by the macro tier. This requires macro and micro tiers to have some coverage overlap so that the UEs can be offloaded. Note that even if a UE associates with the cell providing the highest SIR, it does not mean that this UE will have SIR $> T$ from this cell. As shown in Appendix A, when $K = 1$ and $T \geq 0$ dB, a UE cannot simultaneously be in the coverage area of two different cells \cite{AnalysisMaxSIR}; in these circumstances, there is no overlap area among cells, so that a UE cannot be offloaded from one cell to another. With $K = 1$ and $T < 0$ dB, there is some coverage overlap among cells, however, we will neglect it to carry out the  analysis in Appendix A. On the other hand, if $K > 1$, the coverage overlap can exist for any SIR threshold $T$ value, as considered in sub-Section \ref{subsec:OutageProb}. In this case, our prioritized SIR-based cell association scheme can allow a better load balancing between macro and micro tiers. \myhl{Even though we assume an interference-limited HetNet scenario, the previous considerations are still correct when background noise is included.}

\begin{prop}
\label{proprosionAverageLoad}
\textit{  When prioritized SIR-based cell association with reuse of $K$ is adopted, the probability that a UE associates with the micro tier under the condition that this UE is in the coverage area can be expressed as
}  \begin{equation}\label{eq:microLoad}
    A_{\mu} = \frac{1-(1- {P}_{c,1,\mu})^K}{1-(1- {P}_{c,1})^K},
  \end{equation}
  where $ {P}_{c,1,\mu} = \frac{\lambda_{\mu}\pi P_{\mu}^{2/\gamma} T^{-2/\gamma}}{C(\gamma) \sum_{i = \{M,\mu\}} \lambda_i P_i^{2/\gamma} }$ is the coverage probability of the micro tier when only cells (both macro and micro) using frequency $F_1$ are considered.
\end{prop}

The proof of Proposition \ref{proprosionAverageLoad} is provided in Appendix B, which is adapted from  \cite{AnalysisMaxSIR} taking our prioritized SIR-based cell association scheme into account.

According to our prioritized SIR-based cell association scheme, UEs in the coverage area of the network, but not in the coverage area of the micro tier will associate with the macro tier. Thus, the probability that a reference UE associates with the macro tier (i.e., average load on the macro tier under the condition that the UE is in the coverage area) is complementary with respect to (\ref{eq:microLoad}) as shown below:

\begin{equation}\label{eq:macroLoad}
     A_M  = 1-A_{\mu}
       = \frac{(1-{P}_{c,1,\mu})^K - (1-P_{c,1})^K}{1-(1-P_{c,1})^K}.
\end{equation}

\vspace{10pt}
\section{\uppercase{Rate Coverage Probability}}
\label{sec:Rate}
The rate coverage probability represents the probability that a reference UE can achieve a bit-rate greater than or equal to a certain minimum value denoted by $R_T > 0$ (\footnote{$R_T$ and SIR threshold $T$ correspond to two different constraints: the set of UEs that have bit-rate higher than $R_T$ is a subset of the UEs in the coverage area (i.e., SIR $> T$), assuming that $R_T$ is bigger than the bit-rate corresponding to the minimum SIR = $T$.}) \cite{RateCoverate}. We consider $R_T$ as a given system value. Let $R$ denote the bit-rate (random value) of a reference UE in the network.
When this UE is in outage conditions (not under the coverage area of any tier), its bit-rate is 0 so that the probability that this UE achieves the target rate threshold is 0 as well. Let $C_M$ ($C_{\mu}$) denote the event that the reference UE associates with the macro (micro) tier according to our prioritized SIR-based cell association scheme. A UE is associated with either the macro tier (event $C_M$) or the micro cell tier (event $C_{\mu}$) or it is in outage (event $C_O$). These events are disjoint. Moreover, $\bar{C}_M$ and $\bar{C}_{\mu}$ denote the complementary events of $C_M$ and $C_{\mu}$, respectively. We notice that $C_O = \bar{C}_M \cap \bar{C}_{\mu}$. Then, following the law of total probability, the probability of the event $\{R \geq R_T\}$ can be written as follows:
\begin{equation}\label{eq:rateCover}
\begin{split}
    {P}(R \geq R_T) & =  {P} (R \geq R_T|C_M)  {P}(C_M)  \\
    & +  {P}(R \geq R_T|C_{\mu})  {P}(C_{\mu}) \\
    & +  {P}(R \geq R_T | \bar{C}_M \cap \bar{C}_{\mu})  {P} (\bar{C}_M \cap \bar{C}_{\mu}) \\
     & =  {P}(R \geq R_T|C_M)  {P}(C_M) + \\
     & +  {P}(R \geq R_T|C_{\mu})  {P}(C_{\mu}),
\end{split}
\end{equation}
because $ {P}(R \geq R_T | \bar{C}_M \cap \bar{C}_{\mu}) = 0$ since the UEs in the outage area have no throughput. By using the Bayes rule, we obtain:
\begin{equation}\label{eq:rateCoverBayes}
\begin{split}
    {P}(R \geq R_T) 
     & =  {P}(R \geq R_T, C_M) +  {P}(R \geq R_T, C_{\mu}).
\end{split}
\end{equation}

In order to calculate the rate coverage probability, we will separately derive the rate coverage probabilities of macro and micro tiers in the next sub-Sections.

\subsection{Micro UEs' Rate Coverage Probability}

In this sub-Section, we compute the term $ {P}(R \geq R_T, C_{\mu})$. The event $C_{\mu}$ tells us that the UE, belonging to the micro tier, associates with the micro cell providing the best SIR. We assume that a \emph{Round Robin} (RR) scheduler is used to allocate resources to UEs within a cell. In particular, we divide the bandwidth assigned to a micro cell by the average number of UEs in the micro cell in order to obtain the average bandwidth available per UE in the micro cell. Thus, \myhl {given that the reference UE associates to the micro tier, its bit-rate can be expressed as follows according to the Shannon capacity formula:}
\begin{equation}\label{eq:microRate}
  R  = \frac{{W/K}}{{A_\mu  P_c \;\frac{{\lambda _u }}{{\lambda _\mu  }}}} \log_2 \left(1+ \max_{j \in \Phi_{\mu}} \V{SIR}_j \right),
\end{equation}
where $A_\mu  P_c \frac{{\lambda _u }}{{\lambda _\mu  }}$ denotes the average number of UEs per micro-cell.

Let $\rho_{\mu} = 2^{\frac{R_T K A_{\mu} \lambda_u  {P}_c}{W \lambda_{\mu}}} - 1$. We have:
\begin{equation}\label{eq:microRateCover}
\begin{split}
   & {P}(R \geq R_T, C_{\mu})  =   \\
   & =  {P} \left( \frac{W\lambda_{\mu}}{K A_{\mu}\lambda_u  {P}_c} \log_2 \left(1+ \max_{j \in \Phi_{\mu}} \V{SIR}_j \right) \geq R_T , C_{\mu}\right) \\
   & =  {P} \left(\max_{j \in \Phi_{\mu}} \V{SIR}_j \geq \rho_{\mu}, \cup_{j \in \Phi_{\mu}} \left\{\V{SIR}_j \geq T \right\}\right).
\end{split}
\end{equation}

The event $\max_{j \in \Phi_{\mu}} \V{SIR}_j \geq \rho_{\mu}$ is equivalent to $\cup_{j \in \Phi_{\mu}} \left\{\V{SIR}_j \geq \rho_{\mu}\right\}$ since if one of the SIR values from micro cells is higher than $\rho_{\mu}$, the maximum SIR among those values will be higher than $\rho_{\mu}$ as well and vice versa. Note that depending on the values of $K$, $R_T$, $\lambda_{\mu}$, $T$, and $\lambda_u$, $\rho_{\mu}$ can be bigger or smaller than $T$ (i.e., the rate coverage requirement can be more stringent or less stringent than the outage requirement). Let $T_{\mu} = \max(\rho_{\mu},T)$. Thus, (\ref{eq:microRateCover}) can be re-written as follows:

\begin{equation}\label{eq:microRateCoverSemiFinal}
      {P}(R \geq R_T, C_{\mu}) =  {P} \left(\cup_{j \in \Phi_{\mu}} \left\{\V{SIR}_j \geq T_{\mu}\right\}\right).
\end{equation}

Equation (\ref{eq:microRateCoverSemiFinal}) is fairly easy to interpret because if $\rho_{\mu} < T$, then the outage condition is more stringent than the $R_T$ constraint so that all UEs in the micro tier coverage area will have higher bit-rate than $R_T$. Otherwise, if $\rho_{\mu} \geq T$, equation (\ref{eq:microRateCoverSemiFinal}) can be understood as the micro tier's coverage probability when the SIR threshold is raised from $T$ to $\rho_{\mu}$ [see equation (\ref{eq:coverageProb})]. Then, by adopting the same method as that used to obtain (\ref{eq:coverageProb4}), we derive the coverage probability of the micro tier with SIR threshold $T_{\mu}$ as follows:

\begin{equation}\label{eq:microCoverageProb}
   {P}_{c,\mu} (T_{\mu}) = 1 - \left[1- {P}_{c,1,\mu} (T_{\mu})\right]^K,
\end{equation}
where $ {P}_{c,1,\mu} (T_{\mu}) = \frac{\lambda_{\mu}\pi P_{\mu}^{2/\gamma} T_{\mu}^{-2/\gamma}}{C(\gamma) \sum_{i = \{M,\mu\}} \lambda_i P_i^{2/\gamma} }$ is the micro tier coverage probability when only cells using frequency $F_1$ are considered and SIR threshold is $T_{\mu}$. Appendix B provides the details on the derivation of (\ref{eq:microCoverageProb}) .

Finally, from (\ref{eq:microRateCoverSemiFinal}) and (\ref{eq:microCoverageProb}) we obtain the rate coverage probability of the micro tier as a function of $R_T$ (and then $\rho_{\mu}$) as follows:

\begin{equation}\label{eq:microRateCoverFinal}
   {P}(R \geq R_T, C_{\mu}) = 1 - \left[1- {P}_{c,1,\mu} (T_{\mu})\right]^K.
\end{equation}

\subsection{Macro UEs' Rate Coverage Probability}

In this sub-Section, we compute the term $ {P}(R \geq R_T, C_M)$.
We denote $\rho_M = 2^{\frac{R_T K A_M \lambda_u  {P}_c}{W \lambda_M}} - 1$. The event $C_M$ happens when the reference UE experiences SIR lower than $T$ for all micro cells, but it has SIR higher than $T$ for at least one macro cell. Thus, $C_M$ is characterized by two joint events as follows:
\begin{equation}\label{eq:macroCoverage}
  C_M = \left(\cup_{i \in \Phi_{M}} \left\{\V{SIR}_i \geq T\right\}) \cap (\cap_{j \in \Phi_{\mu}} \left\{\V{SIR}_j < T\right\}\right).
\end{equation}

Similar to the previous part, the rate coverage probability of the macro tier can be re-written as follows:
\begin{equation}\label{eq:macroRateCover}
  \begin{split}
     & {P}(R \geq R_T, C_M) \\
       & =   {P}\left(\max_{i \in \Phi_{M}} \V{SIR}_i \geq \rho_M, \cup_{i \in \Phi_{M}} \{\V{SIR}_i \geq T\},\cap_{j \in \Phi_{\mu}} \{\V{SIR}_j < T\} \right)\\
       & =  {P}\left(\cup_{i \in \Phi_{M}} \left\{\V{SIR}_i \geq \max(\rho_M,T)\right\}, \cap_{j \in \Phi_{\mu}} \{\V{SIR}_j < T\} \right).
  \end{split}
\end{equation}
In order to write the above, we have exploited the fact that $\max_{i \in \Phi_{M}} \V{SIR}_i \geq \rho_M$ is equivalent to $\cup_{i \in \Phi_{M}} \left\{\V{SIR}_i \geq \rho_M \right\}$, as in the previous case. Let $T_M = \max(\rho_M,T)$ and $D_M$ be the event $\cup_{i \in \Phi_{M}} \{\V{SIR}_i \geq T_M\}$. Note that the event $\cap_{j \in \Phi_{\mu}} \{\V{SIR}_j < T\}$ is exactly $\bar{C}_{\mu}$. We consider the following relation among events and corresponding probabilities:

\begin{equation}\label{eq:setToEvent}
 \begin{split}
   & D_M \cup C_{\mu} = C_{\mu} \cup D_M \cap \bar{C}_{\mu}
     \Rightarrow \\ & \quad \quad {P} (D_M \cup C_{\mu}) =  {P} (C_{\mu} \cup D_M \cap \bar{C}_{\mu}).
  \end{split}
\end{equation}
Since $C_{\mu} $ and $D_M \cap \bar{C}_{\mu}$ are disjoint events, we have:
\begin{equation}\label{eq:eventProb}
   {P} (D_M \cup C_{\mu}) =  {P}(C_{\mu}) +  {P}(D_M \cap \bar{C}_{\mu}).
\end{equation}
From (\ref{eq:macroRateCover}) and (\ref{eq:eventProb}), we have:
\begin{equation}\label{eq:macroRateCoverFinal}
  \begin{split}
    &  {P}(R \geq R_T, C_M) \\
    & =  {P}\left(  D_M \cap \bar{C}_{\mu} \right)
        =   {P} \left(D_M \cup C_{\mu}\right) -   {P} (C_{\mu})\\
       & =   {P} \left( (\cup_{i \in \Phi_{M}} \{\V{SIR}_i \geq T_M\}) \cup (\cup_{j \in \Phi_{\mu}} \{\V{SIR}_j \geq T\}) \right)  \\
       & \quad \quad -  {P} \left(\cup_{j \in \Phi_{\mu}} \{\V{SIR}_j \geq T\} \right) \\
       & = 1 - \left[ 1 -  {P}_{c,1}(T_M,T)\right]^K - 1 + \left[ 1- {P}_{c,1,\mu} (T) \right]^K \\
       & = \left[1- {P}_{c,1,\mu}(T) \right]^K - \left[ 1 -  {P}_{c,1}(T_M,T)\right]^K,
  \end{split}
\end{equation}
where $ {P} \left( (\cup_{i \in \Phi_{M}} \{\V{SIR}_i \geq T_M\}) \cup (\cup_{j \in \Phi_{\mu}} \{\V{SIR}_j \geq T\}) \right)$ has been derived in Appendix A, referring only to the cells using $F_1$; then, we generalize to all frequency segments analogously to what is shown in (\ref{eq:coverageProb4}). Still in Appendix A, we show in (\ref{th:noiseIgnored}) that $ {P}_{c,1}(T_M,T) = \frac{\pi}{C(\gamma)}\frac{\lambda_M P_M^{2/\gamma} T_M^{-2/\gamma} + \lambda_{\mu} P_{\mu}^{2/\gamma}T^{-2/\gamma}}{\sum_{i = \{M,\mu\}} \lambda_i P_i^{2/\gamma}}$, the coverage probability of the network (including both macro and micro tiers) when only the sub-set of cells using frequency $F_1$ are considered and the SIR threshold of the macro tier is $T_M = \max(\rho_M,T)$.

\subsection{Combination of the Two Cases and Mean Bit-Rate}

From (\ref{eq:rateCoverBayes}), (\ref{eq:microRateCoverFinal}), and (\ref{eq:macroRateCoverFinal}), we obtain the rate coverage probability as follows:
\begin{equation}\label{eq:rateCoverFinal}
\begin{split}
     {P}(R \geq R_T) = & 1 - \left[1- {P}_{c,1,\mu} (T_{\mu})\right]^K + \left[1- {P}_{c,1,\mu} (T) \right]^K \\
     & - \left[ 1 -  {P}_{c,1}(T_M,T)\right]^K.
\end{split}
\end{equation}
Further elaborating (\ref{eq:rateCoverFinal}) by means of the Netwon Bimomial formula, we obtain the expression in (\ref{eq:rateCoverFinal2}) that is shown at the top of the next page, where the only terms depending on $R_T$ are $T_\mu$ and $T_M$. This expression is useful in what follows for performing the integration over $R_T$.
\begin{figure*}
\begin{equation}\label{eq:rateCoverFinal2}
\begin{split}
{P}(R \geq R_T) & =  \sum\limits_{k = 1}^K {\left( { - 1} \right)^{k + 1} \binom{K}{k} \left\{ {\left[ {P_{c,1,\mu }^{} \left( {T_\mu  } \right)} \right]^k  - \left[ {P_{c,1,\mu }^{} \left( T \right)} \right]^k  + \left[ {P_{c,1}^{} \left( {T_M ,T  } \right)} \right]^k } \right\}} \\
&   = \sum\limits_{k = 1}^K {\left( { - 1} \right)^{k + 1} \binom{K}{k} D^k(\gamma, T)}  \frac{{\left( {\frac{{T_\mu  }}{T}} \right)^{\frac{{ - 2k}}{\gamma }}  + \sum\limits_{i = 1}^k {\left( {\begin{array}{*{20}c}
   k  \\
   i  \\
\end{array}} \right)\left[ {\left( {\frac{{\lambda _M }}{{\lambda _\mu  }}} \right)\left( {\frac{{P_M }}{{P_\mu  }}} \right)^{\frac{2}{\gamma }} \left( {\frac{{T_M }}{T}} \right)^{\frac{{ - 2}}{\gamma }} } \right]^i } }}{{\left[ {1 + \left( {\frac{{\lambda _M }}{{\lambda _\mu  }}} \right)\left( {\frac{{P_M }}{{P_\mu  }}} \right)^{\frac{2}{\gamma }} } \right]^k }}.
\end{split}
\end{equation}
\end{figure*}

We expect that $ {P} (R \geq R_T)$ increases as $R_T$ decreases to 0. When $R_T$ is sufficiently small so that $\max(\rho_M,\rho_{\mu}) \leq T$, we have $ {P}_{c,1,\mu} (T_{\mu}) \equiv  {P}_{c,1,\mu} (T)$, $ {P}_{c,1}(T_M,T) \equiv  {P}_{c,1}(T)$ and the following rate coverage probability formula:

\begin{equation}\label{eq:rateSubCase}
   {P}(R \geq R_T) \equiv 1  - \left( 1 -  {P}_{c,1} (T) \right)^K,
\end{equation}
which is exactly the coverage probability of the network, $ {P}_c$. In this case, all UEs in the coverage area have bit-rate satisfying the rate threshold. Thus, the maximum value of the rate coverage probability is equal to the coverage probability of the network.

We can also consider the average of the UE bit-rate, $R$, that represents an important parameter characterizing the performance provided to a UE. We can compute this by using the distribution corresponding to the rate coverage probability as follows:
\begin{equation}\label{average_UE_ratel}
E\left[ R \right] = \int_0^{ + \infty } {P \left\{ {R > R_T } \right\}dR_T }.
\end{equation}

The mean UE bit-rate can be obtained by applying the above integral to the rate coverage probability expression in (\ref{eq:rateCoverFinal2}); then, by exploiting the linearity of the integral operator, we resort to apply the integral to two types of terms, as detailed below:
\begin{equation}\label{average_UE_ratel2}
\begin{array}{l}
 \int_0^{ + \infty } {\left[ {T_\mu ^{\frac{{ - 2k}}{\gamma }} } \right] \V{d}R_T   = \quad } \int_0^{ + \infty } {\left[ {\max \left\{ {\rho _\mu  ,T} \right\}} \right]^{\frac{{ - 2k}}{\gamma }}} \V{d}R_T \quad   \\
 \int_0^{ + \infty } {\left[ {T_M^{\frac{{ - 2k}}{\gamma }} } \right] \V{d}R_T  = \int_0^{ + \infty } {\left[ {\max \left\{ {\rho _M ,T} \right\}} \right]^{\frac{{ - 2k}}{\gamma }} \V{d}R_T \quad } } . \\
 \end{array}
\end{equation}

\begin{figure*} [!h]
\begin{equation}\label{average_UE_ratel3}
\int_0^{ + \infty } {\left[ {\max \left\{ {\rho _{} ,T} \right\}} \right]^{ - b} \V{d}R_T \quad }  = \int\limits_0^{\frac{1}{a}\log _2 \left( {1 + T} \right)} {\frac{1}{{T^b }} \V{d}R_T }  + \int\limits_{\frac{1}{a}\log _2 \left( {1 + T} \right)}^{ + \infty } {\frac{1}{{\left( {2^{aR_T }  - 1} \right)^b }} \V{d}R_T }  = \frac{1}{{aT^b }}\log _2 \left( {1 + T} \right) + \frac{{B_{\frac{1}{{1 + T}}} \left( {b,\;1 - b} \right)}}{{a\ln \left( 2 \right)}}.
\end{equation}
\end{figure*}

These integrals can be expressed by means of the incomplete Beta function $B_x(\alpha,\beta)$ \cite{betafunction} as shown in (\ref{average_UE_ratel3}) at the top of the next page,
 where $\rho = \rho_{\{\mu ~{\rm  or }~M\}}$, $a = a_{\{\mu ~{\rm  or }~M\}}  = \frac{{KA_{\left\{ {\mu \;{\rm or}\;M} \right\}} P_c \lambda _u }}{{W\lambda _{\left\{ {\mu \;{\rm or}\;M} \right\}} }}$, and $b = 2k/\gamma$ ($k$ is here a generic integer value from 1 to $K$). The minimum $R_T$ value (i.e., the value corresponding to the SIR threshold value $T$) for the macro cell coverage is $\frac{1}{{a_M  }}\log _2 \left( {1 + T} \right)$ and the minimum $R_T$ value for the micro cell coverage is $\frac{1}{{a_\mu  }}\log _2 \left( {1 + T} \right)$. However, in equations (\ref{eq:rateCoverFinal}) and (\ref{eq:rateCoverFinal2}), $R_T$ can also be below the previous minimum bit-rate values (so that the integral in (\ref{average_UE_ratel}) starts from $R_T = 0$), because in these circumstances, the rate coverage probability coincides with the coverage probability and its value is independent of $R_T$. Finally, numerical methods have to be used to compute (\ref{average_UE_ratel3}) to determine the mean UE bit-rate.

Note that the approach in (\ref{average_UE_ratel}) to determine the mean UE bit-rate is different from that adopted in \cite{AnalysisMaxSIR}, because in that work the authors refer to a mean UE bit-rate (not considering that there are many UEs that share the cell capacity as we do here introducing coefficients $a_{\mu}$ and $a_M$) conditioned on the coverage and referring to a simple max-SINR cell association scheme with no frequency reuse.


In this analysis of outage, cell load, and rate coverage probabilities, basic parameters are: $\gamma$, $T$, $W$, and $K$. Instead, the other parameters ($\lambda_M$, $\lambda_\mu$, $P_M$, $P_\mu$, $\lambda_u$) influence the numerical results only via the following ratios: $\lambda_M / \lambda_\mu$, $P_M / P_\mu$, and $\lambda_u / \lambda_M$. Thus, in Section \ref{sec:ResultsPPP}, we present an optimization approach for both $K$ and $\lambda_{\mu}/\lambda_M$ (given the other parameters). Hence, using the obtained model and analysis, network operators can select both $K$ and  and decide when it is convenient to increase the density of $\mu$-eNBs ($\lambda_{\mu}/\lambda_M$) in the system, so that users' quality of experience is satisfied (see Figs. 15 and 16 in Section \ref{sec:ResultsPPP}).

Even if we consider a two-tier HetNet system, this work could be extended to more than two tiers. In particular, we could add a femto layer, where each cell selects a frequency segment at random as well. Then, we can apply our prioritized SIR-based cell association scheme, using a priority order for cell associations as femto $>$ micro $>$ macro. The detailed scheme with more than two tiers is left to a future study.

\vspace{10pt}
\section{\uppercase{Simulation Approach and Settings}}
\label{sec:Simulation}
It is important to notice that the simulation of a PPP-based cellular system is quite different from that of a hexagonal-regular system, but the main idea is the same. In the case of a hexagonal-regular cellular network, a simulation is basically organized with a central macro cell and surrounding interfering cells and we extract performance results only from the central macro cell to avoid border effects. This is possible since all the cells have the same shape and size so that the central cell is taken as representative of the whole network. This method, however, is no longer applicable with stochastic geometry, where cells have different shapes so that it is difficult to identify a central cell.

In the case of stochastic geometry, we have to take a UE as a reference (not a cell as a reference) in the origin around which all cells and all other UEs are scattered according to PPPs. Moreover, in the analysis, the area where PPP is applied is an infinite plane, so that all interference sources are taken into account and border effects are not present. In the simulation, we implemented PPP on a very large area with the reference UE at the center so that the interference from cells outside the area to the UE is negligible and border effects are eliminated.  All macro and micro eNBs are placed on a plane with sizes $L \times L$, where $L$ is the length of the side of the plane (thus, the average number of macro cells is $N_c = \lambda_{M}L^2$, the average number of micro cells is $N_r = \lambda_{\mu}L^2$, and the average number of UEs is $N_u = \lambda_{u}L^2$). To generate points according to PPPs, we first determine the number of points in $L^2$ using Poisson random variables with mean values $N_c$, $N_r$, and $N_u$. Then, conditioning on the number of points in the $L^2$ area of the PPPs of M-eNBs, \textmu-eNBs, and UEs, the position of each point is determined according to a uniform distribution in $L^2$.

Each simulation is repeated $N = 10000$ times, regenerating at each run the positions of M-eNBs, \textmu-eNBs, and UEs according to the corresponding homogeneous PPP processes. In each simulation, depending on the topology, the reference UE can associate with macro or micro tier or can be in the outage area. By repeating the simulation $N$ times and observing the results, we can obtain the probability that the UE is in the coverage area or in outage conditions, and the probability that the UE belongs to the macro tier or the micro tier.

The LTE-A HetNet scenario has been implemented in a Matlab simulator, using the Monte Carlo approach. In details, we simulate a square area with side $L = 20$ km and an average number of 80 M-eNBs ($\lambda_{M} = \frac{80}{400} = 0.2 $ $\V{M-eNBs/km}^2$). The density of UEs is $\lambda_u = 100 \lambda_{M}$. The transmission power of M- (\textmu-) eNBs is 46 dBm (30 dBm). The system bandwidth is $W$ = 20 MHz. If not differently stated, SIR threshold $T$ is set to $0$ dB, micro cells density is $\lambda_{\mu} = 4\lambda_{M}$, pathloss exponent $\gamma$ is 4, and rate threshold $R_T$ is set to 1 Mbps.

\vspace{10pt}
\section{\uppercase{Results}}
\label{sec:ResultsPPP}
In this Section, we verify our analysis via simulations. Moreover, we compare our prioritized SIR-based cell association scheme with other schemes in the literature, which are: max-\emph{Reference Signal Received Power} (RSRP)\footnote{In the LTE-A standard, RSRP is defined as the linear average over the power contributions of the \emph{Resource Elements} (REs) that carry cell-specific reference signals within the considered measurement frequency bandwidth \cite{TS36211}.} with spectrum sharing \cite{HetNEtFlexibleAssoc},\cite{OffloadinginHetNets}, max-biased RSRP with spectrum partitioning \cite{SpectrumAllocInfocom},\cite{AUtilityPerspective},\cite{RateCoverate},\cite{AnalysiswithCellUnderLoad}, and max-SIR with frequency reuse of $K$. Our aim is to show the advantages of our prioritized SIR-based scheme in terms of outage probability, load balancing, and rate coverage probability. We also validate the analysis of the mean UE bit-rate by means of simulations. Finally, We propose an optimization approach to select $K$ and $\lambda_{\mu}/\lambda_M$ values to satisfy certain cell planning requirements.

\subsection{Model and Analysis Validation}

\begin{figure}
  \centering \epsfxsize=8.5cm \epsfbox{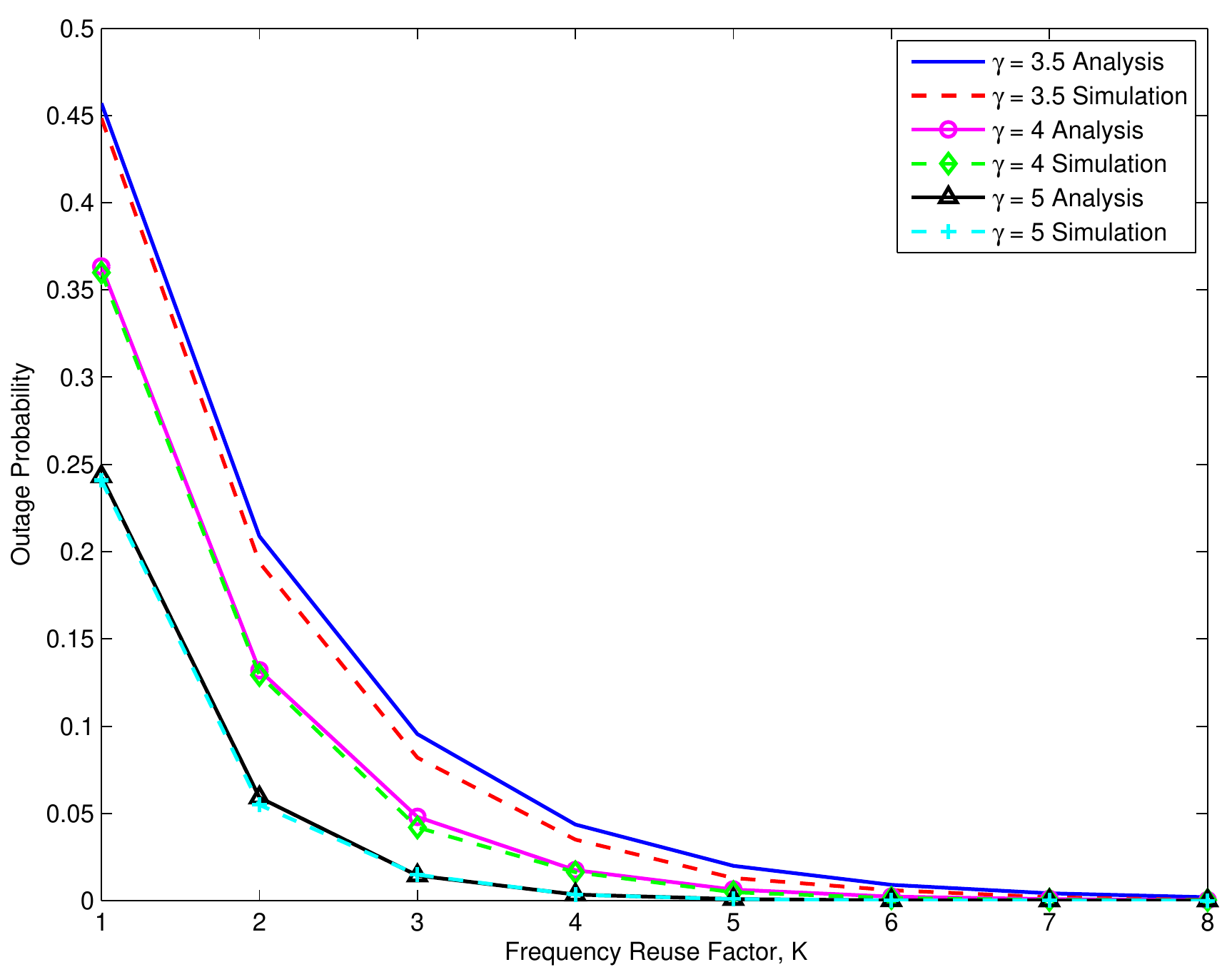}
  \caption{Outage probability as function of frequency reuse factor $K$.}\label{fig:OutagevsKGamma}
\end{figure}

\begin{figure}
  \centering \epsfxsize=8.2cm \epsfbox{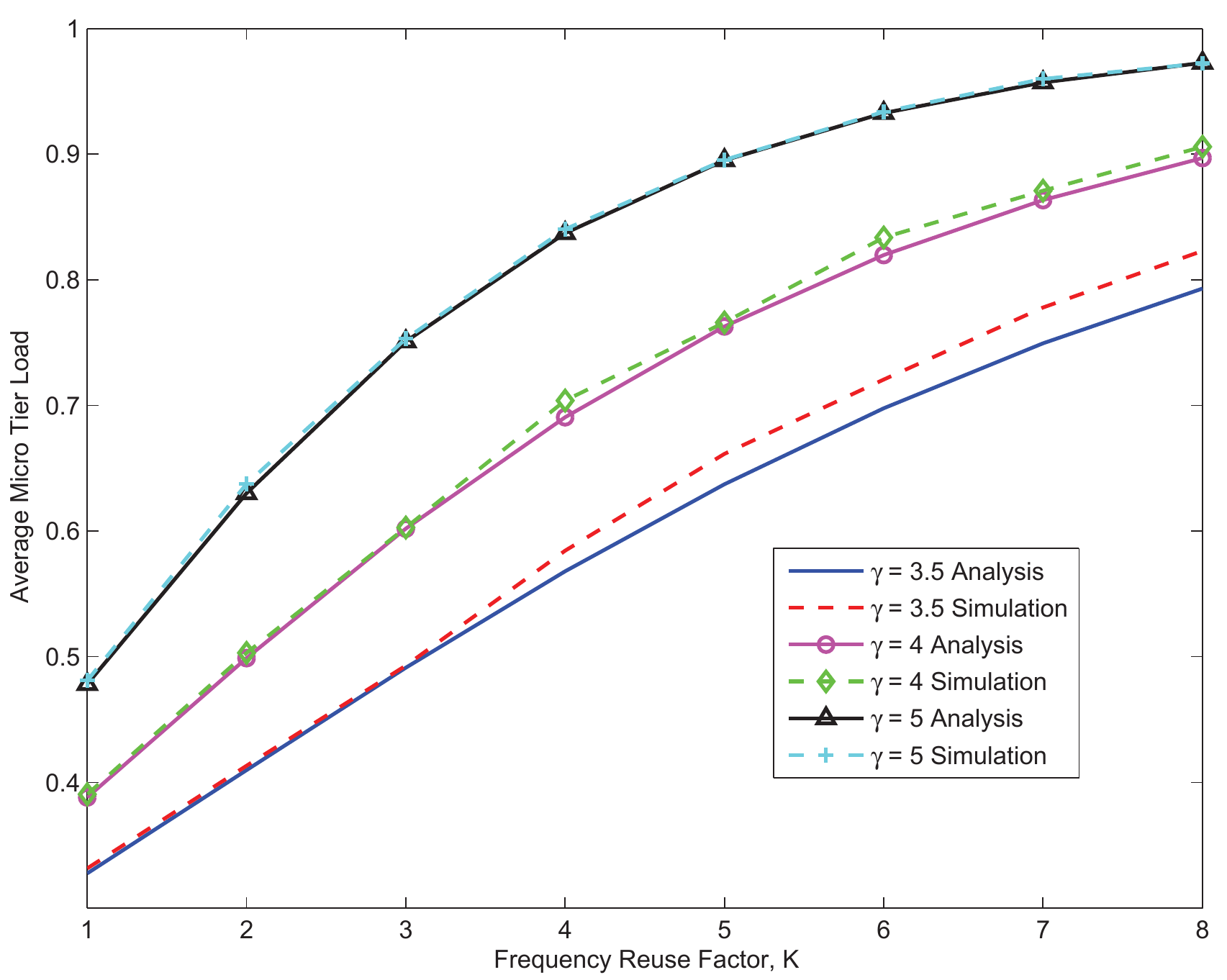}
  \caption{Average load on the micro tier as function of frequency reuse factor $K$.}\label{fig:LoadvsKGamma}
\end{figure}

\begin{figure}
  \centering \epsfxsize=8cm \epsfbox{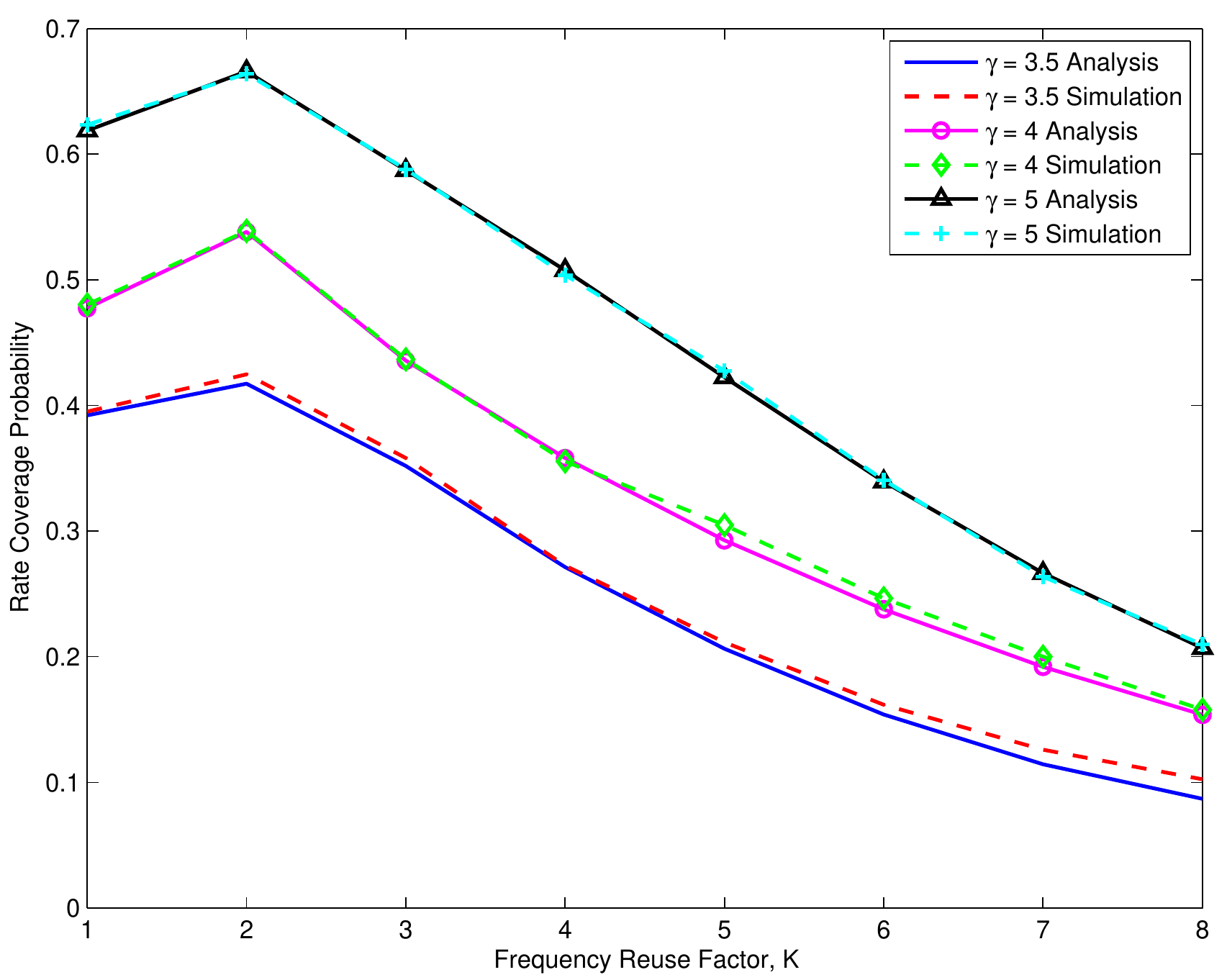}
  \caption{Rate coverage probability as function of frequency reuse factor $K$.}\label{fig:RatevsKGamma}
\end{figure}

Let us focus on the first set of results (Figs. \ref{fig:OutagevsKGamma}, \ref{fig:LoadvsKGamma}, and \ref{fig:RatevsKGamma}), where we vary the frequency reuse factor $K$ from 1 to 8. Note that $K = 1$ means that every cell uses the entire spectrum, which is equivalent to the spectrum sharing scheme proposed in \cite{AnalysisMaxSIR}. The path loss exponent can assume different values such as 3.5, 4, or 5, which are compatible with a urban environment. We can see from Fig. \ref{fig:OutagevsKGamma} that the outage probability rapidly reduces with $K$. In particular, the outage probability for $\gamma = 4$ decreases from 36\% when $K = 1$ to 13\% when $K = 2$ and to 5\% when $K = 3$; with $K \geq 5$, there is almost no UEs in outage conditions. Fig. \ref{fig:LoadvsKGamma} shows that the average load on the micro tier $A_{\mu}$ increases from 38\% to nearly 60\% when $K$ increases from 1 to 3. The reason is that the micro tier can provide better SIR to UEs  with larger $K$ value so that more UEs associate with the micro tier by means of the prioritized SIR-based cell association scheme. The rate coverage probability is shown in Fig. \ref{fig:RatevsKGamma}. It can be observed that $K=2$ gives the best results in this case ($\lambda_{\mu} = 4 \lambda_M$) for all the $\gamma$ cases. Beyond this $K$ value, the rate coverage probability reduces since the bandwidth share of each UE is smaller. The optimal value of $K$ will change when the ratio between micro cells density and macro cell density changes, as discussed later in this Section. When the path loss exponent is higher, we achieve better performance. The reason is that in this dense HetNet scenario, the high path loss exponent makes each cell to become more isolated from the rest of the network, thus reducing the interference among cells. As a final remark, we can see that the analysis results are very close to simulation ones, thus validating our theoretical approach.

\begin{figure}
   \centering \epsfxsize=8.5cm \epsfbox{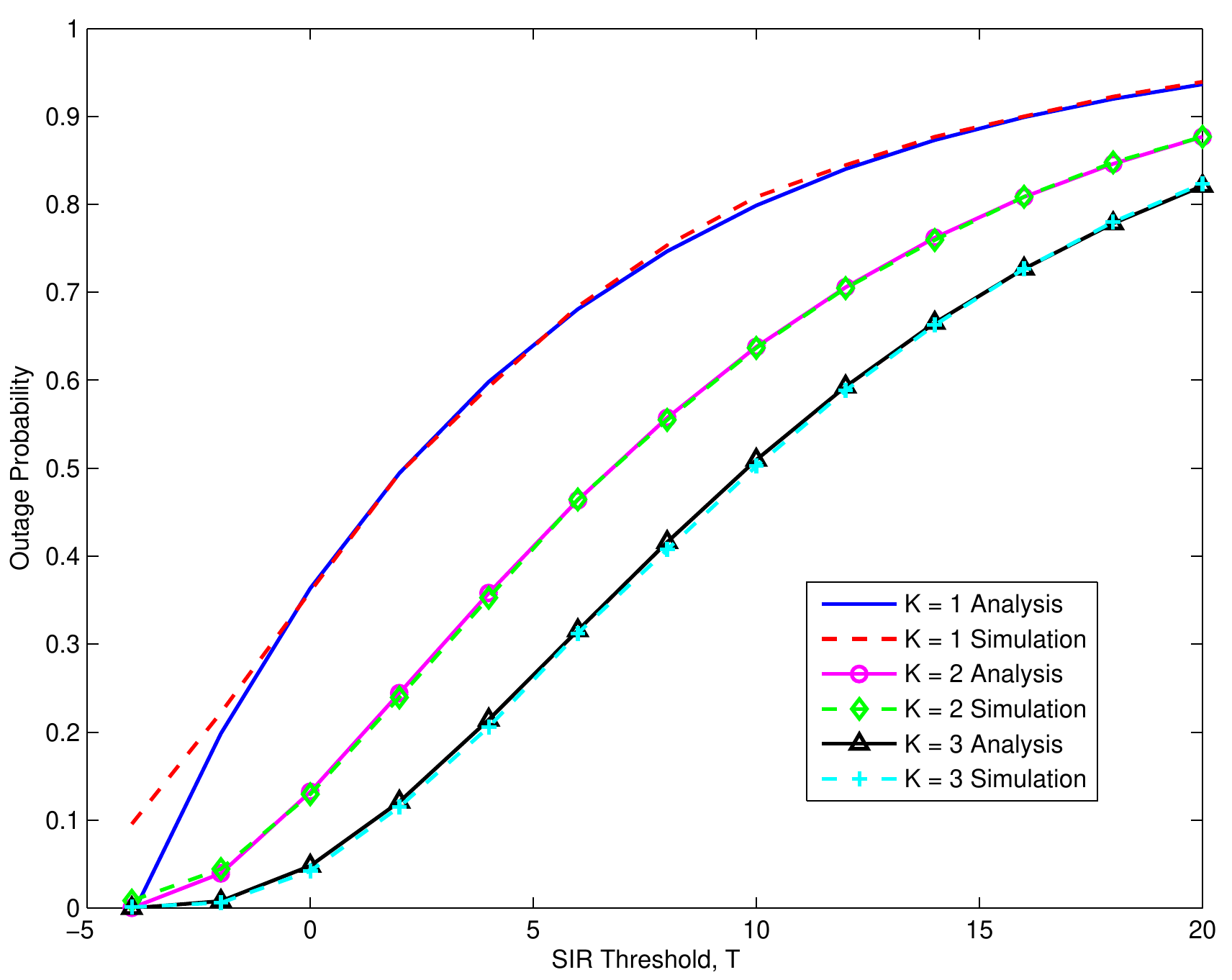}
   \caption{Outage probability as function of SIR threshold $T$ in dB.}\label{fig:OutagevsSIR}
\end{figure}

\begin{figure}
  \centering \epsfxsize=8.5cm \epsfbox{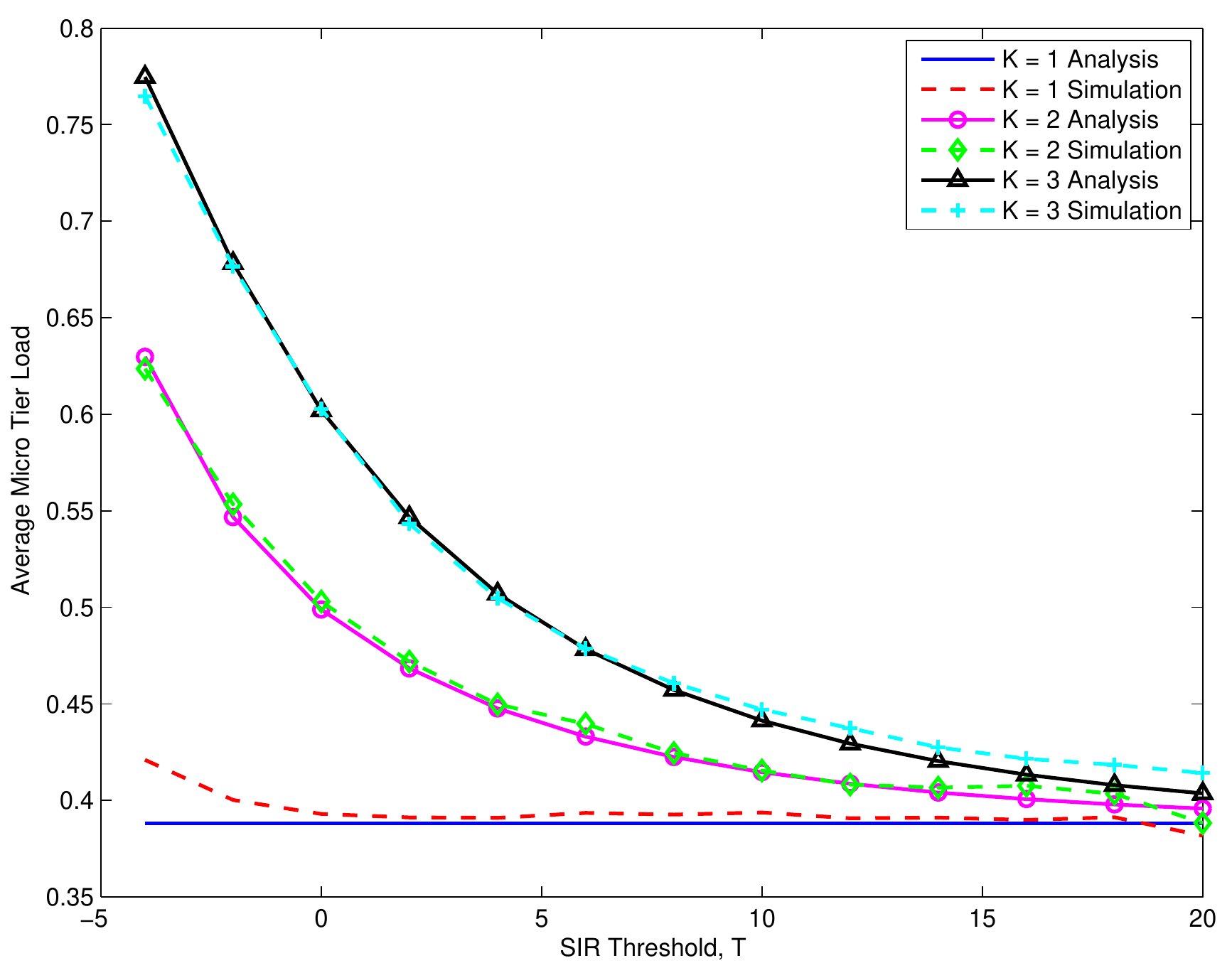}
  \caption{Average load on the micro tier as function of SIR threshold $T$ in dB.}\label{fig:LoadvsSIR}
\end{figure}

\begin{figure}
  \centering \epsfxsize=9.2cm \epsfbox{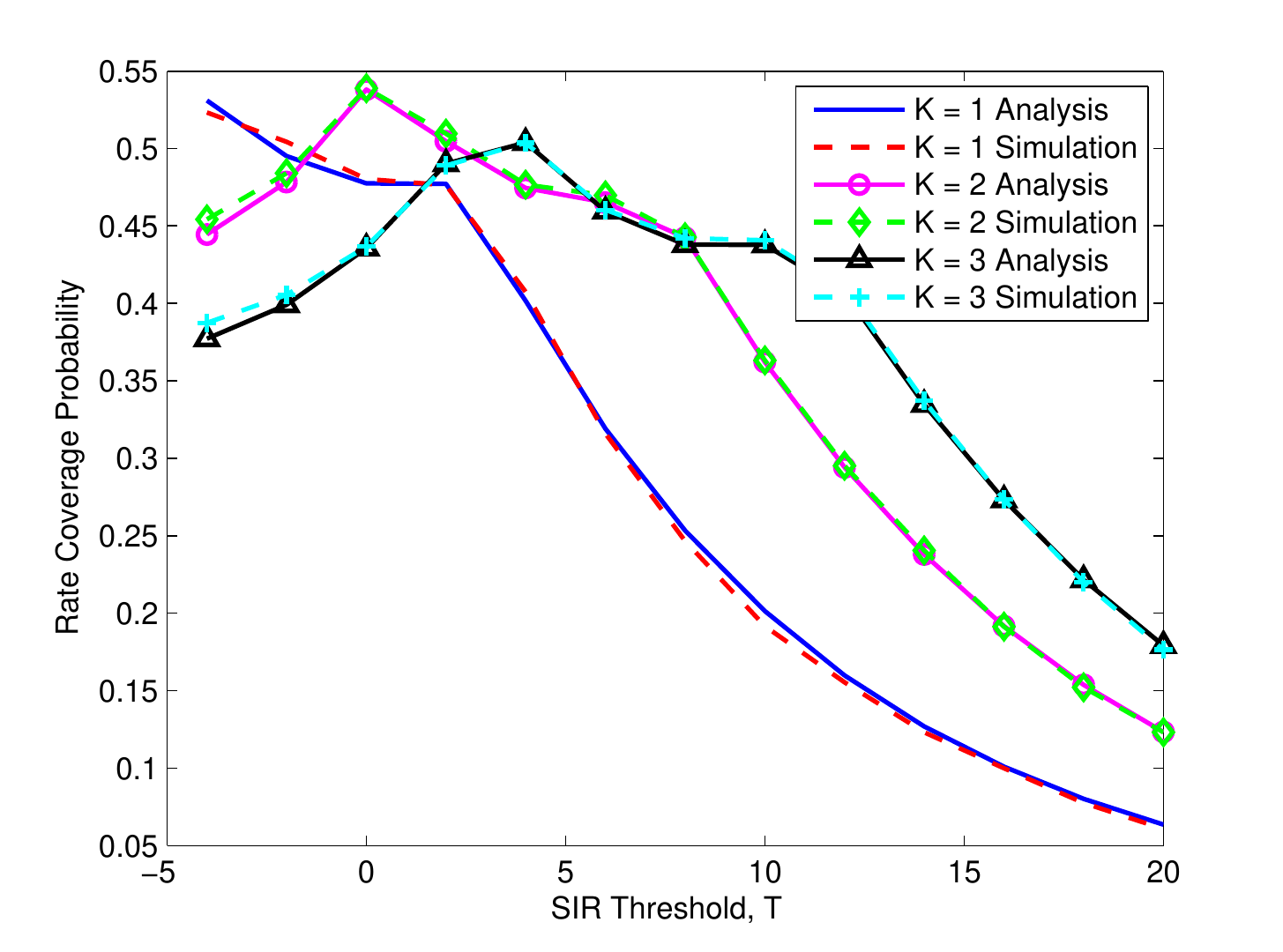}
  \caption{Rate coverage probability as function of SIR threshold $T$ in dB.}\label{fig:RatevsSIR}
\end{figure}

Figures \ref{fig:OutagevsSIR}, \ref{fig:LoadvsSIR}, and \ref{fig:RatevsSIR} show the outage probability, average load, and rate coverage probability for SIR threshold $T$ (both macro tier and micro tier) from $-4$ dB to $20$ dB with different values of $K$. We can see from Fig. \ref{fig:OutagevsSIR} that the outage probability reduces with $K$. When $K = 1$, analytical results closely follow simulation ones for $T \geq -2$ dB; if $T < -2$ dB, however, the analysis provides a lower outage probability as compared with simulation results because of the approximation of disjoint cells in the coverage analysis, as explained in Appendix A [see formula (\ref{th:coverageProb})]. Instead, the analysis provides very close results to simulations for $K > 1$ even if $T < 0$ dB, because the approximation improves with $K$ as stated in the Appendix A. SIR threshold $T$ has also impact on both the average load on the micro tier, $A_{\mu}$, and the rate coverage probability, $ {P} (R \geq R_T)$, as shown in Figs. \ref{fig:LoadvsSIR} and \ref{fig:RatevsSIR}. In particular, with lower $T$ values, there are more UEs under the coverage area, which leads to more UEs associated with the micro tier as a result of our prioritized SIR-based cell association scheme. Thus, SIR threshold $T$ can be used as a parameter to control the average load on each tier. Fig. \ref{fig:LoadvsSIR} also shows that the average load from analysis does not depend on $T$ when $K = 1$, because, as we can see from (\ref{eq:microLoad}), $A_{\mu}$ is independent of $T$. The rate coverage probability has a maximum when $T = 0$ dB and $K = 2$ as shown in Fig. \ref{fig:RatevsSIR}. The reason is that when $T$ increases there are more UEs in outage conditions, thus leading to more resources for UEs which are still under the coverage area; however, the number of UEs in the coverage area reduces, so that the rate coverage probability reduces as well.


\begin{figure}
  \centering \epsfxsize=8cm \epsfbox{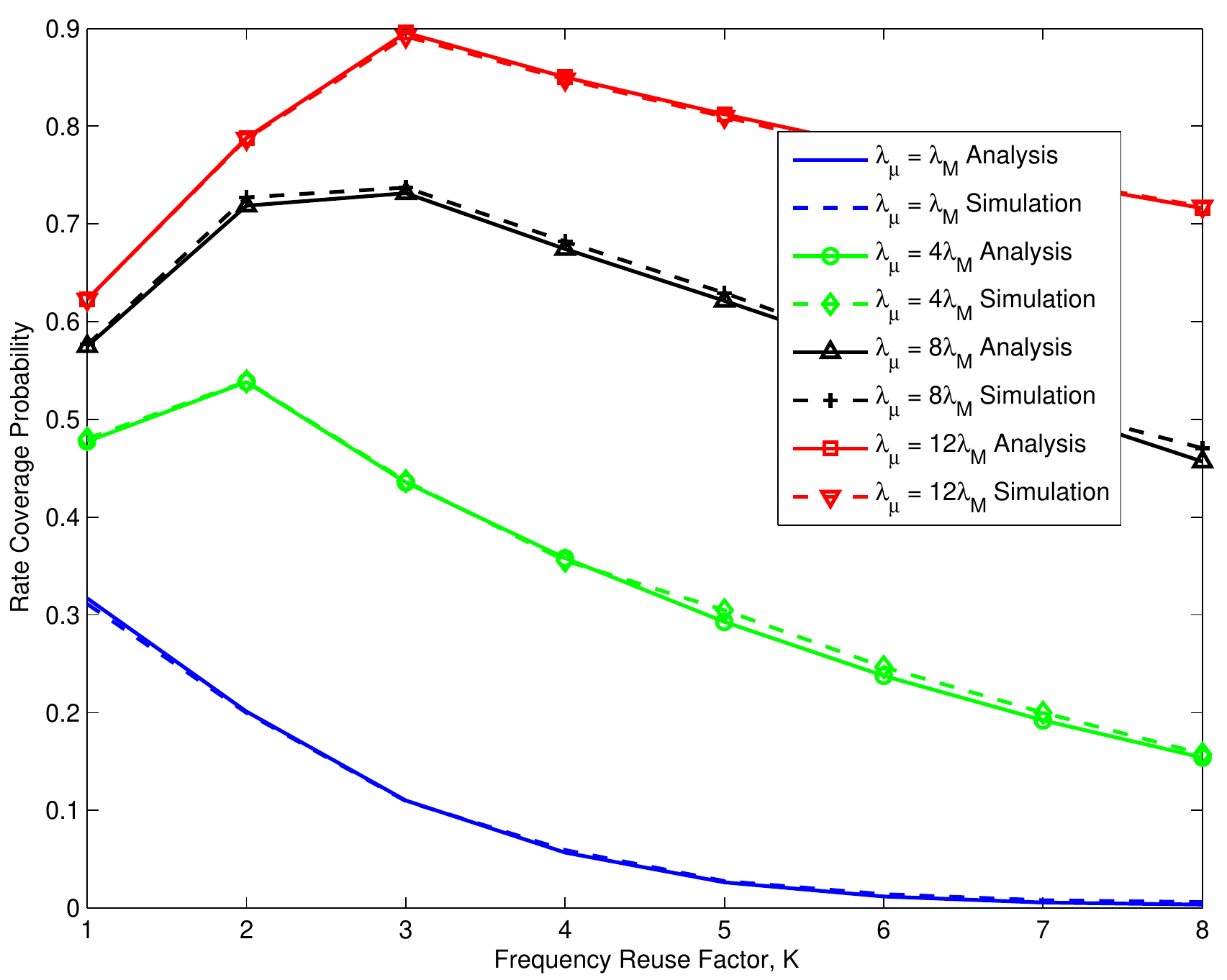}
  \caption{Rate coverage probability with different micro density-to-macro density ratios $\lambda_{\mu}/\lambda_M$.}\label{fig:RatevsRatio}
\end{figure}

Figure
\ref{fig:RatevsRatio} shows that there are different $K$ values maximizing the rate coverage probability for different $\lambda_{\mu}/\lambda_M$ values and $T$ = 0 dB. In fact, when $\lambda_{\mu} = \lambda_M$, we should use $K = 1$. On the other hand, when $\lambda_{\mu} = 4\lambda_M$, it is better to use $K = 2$; instead, $K = 3$ should be adopted when $\lambda_{\mu} = 8\lambda_M$ or higher. This result can be justified because the more the micro cells, the higher the interference in the network, so that bigger $K$ values should be adopted.

\begin{figure}
  \centering \epsfxsize=8cm \epsfbox{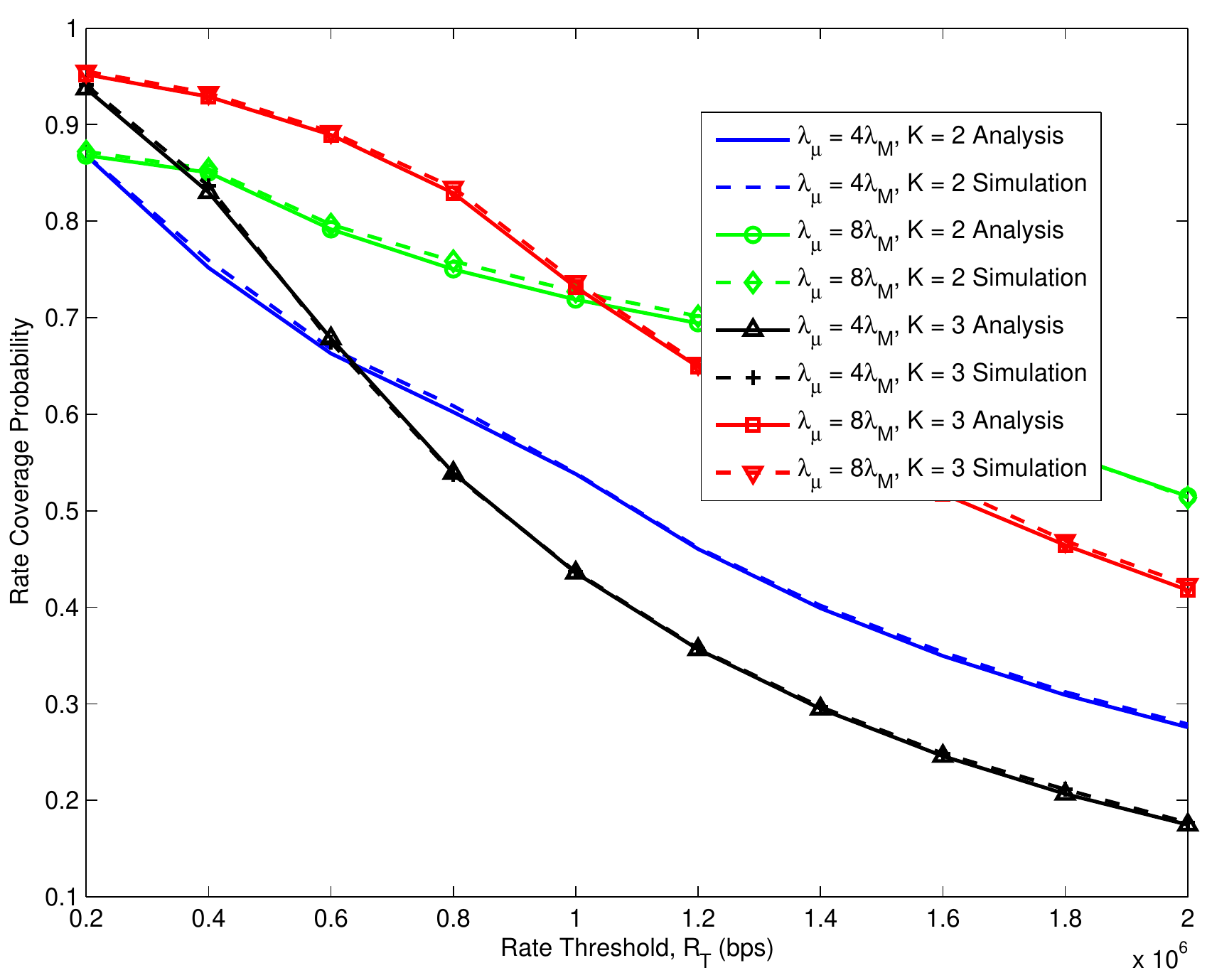}
    \caption{Rate coverage probability as function of rate threshold $R_T$.}\label{fig:RatevsThreshold}
\end{figure}

In Fig. \ref{fig:RatevsThreshold}, we show the rate coverage probability as a function of the rate threshold $R_T$ ranging from 200 kbps to 2 Mbps. We can observe that higher $\lambda_{\mu}/\lambda_M$ values provide better performance for the same $K$. We note that $K = 3$ provides higher rate coverage probability than $K = 2$ for small rate threshold $R_T$. In fact, $K = 3$ can help to better control interference in the network, thus improving bit-rate of edge UEs. Moreover, with $K = 3$, there are more UEs served by the micro tier as shown in Fig. \ref{fig:LoadvsKGamma}, thus traffic is more balanced than with $K = 2$. Instead, $K = 2$ shows better results than $K = 3$ with high rate threshold values $R_T$ since center UEs (having low interference) can benefit from more bandwidth with smaller $K$. Finally, we can note that there is a very close agreement between analysis and simulations in all the cases.

\begin{figure}
  \centering \epsfxsize=8cm \epsfbox{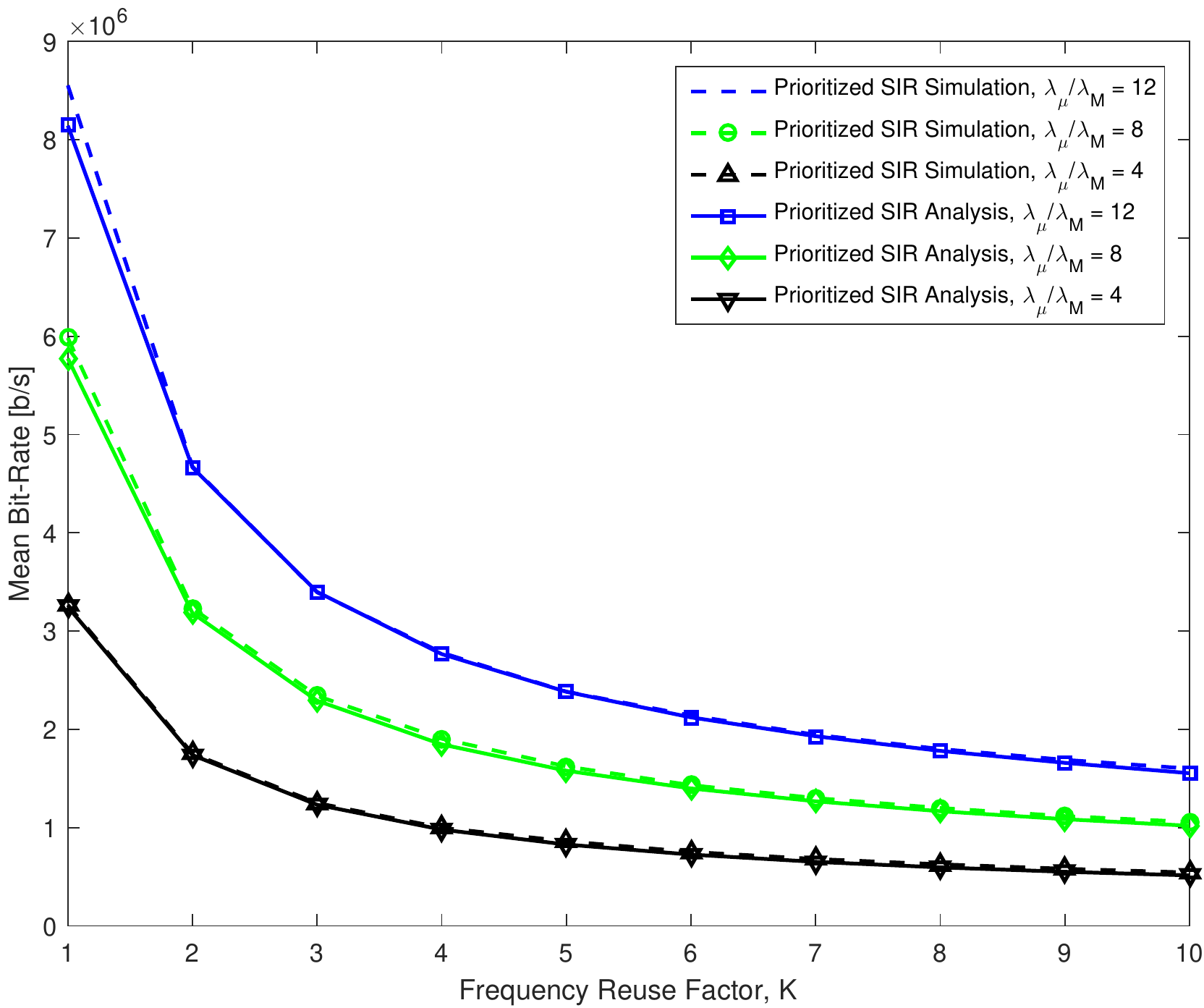}
  \caption{Mean UE bit-rate rate as a function of the frequency reuse $K$.}\label{fig:meanUErate}
\end{figure}

In Fig. \ref{fig:meanUErate}, we show the mean UE bit-rate rate as a function of the frequency reuse factor $K$ for different densities of the micro cells (and given density of macro cells) represented by $\lambda_{\mu}/\lambda_M$ equal to 4, 8, and 12. There is a very good agreement between simulations and the analytical results obtained according to (\ref{average_UE_ratel})-(\ref{average_UE_ratel3}). As expected, the mean UE bit-rate decreases with $K$ and increases with $\lambda_{\mu}/\lambda_M$.

\subsection{Comparisons with Other Schemes}

In this sub-Section, we compare our prioritized SIR-based cell association with other schemes in the literature. In particular, the following schemes are considered for comparisons:\\

\begin{itemize}[leftmargin=.2in]
  \item Max-RSRP cell association with spectrum sharing \cite{HetNEtFlexibleAssoc},\cite{OffloadinginHetNets}: in this scheme, all macro and micro cells share the same frequency band ($K = 1$). A generic UE associates with the cell providing the highest RSRP value.
 \vspace{0.1cm} \item Max-RSRP cell association with $K = 3$: this scheme uses the same frequency reuse technique as in our study, but with an RSRP-based cell association criterion.
  \vspace{0.1cm}\item Max-RSRP cell association with \emph{Resource Partitioning 1} (RP1) \cite{SpectrumAllocInfocom},\cite{AUtilityPerspective}: in this scheme, the frequency band is divided into 2 parts, $F_1$ and $F_2$. All macro cells use $F_1$, while all micro cells use $F_2$. Thus, the two tiers do not interfere with each other.
  \vspace{0.1cm}\item Max-biased RSRP cell association with \emph{Resource Partitioning 2} (RP2): this scheme is used in \cite{AnalysiswithCellUnderLoad} and \cite{RateCoverate}. Radio resources are allocated differently from the previous RP1 scheme as described below, considering three groups of UEs, such as macro UEs, micro UEs, and biased UEs. In particular, macro UEs experience higher received power (i.e., RSRP) from the macro tier even though bias is applied to the micro tier. Micro UEs experience higher received power from the micro tier even when bias is not used at the micro tier. Finally, biased UEs experience the best received power from the macro tier when bias is not applied, but they obtain better received power from the micro tier when bias is applied to them, so that these UEs are forced to associate with the micro tier. Hence, the difference between max-RSRP and max-biased RSRP is in the association of biased UEs (associated with the macro tier with max-RSRP scheme and associated with the micro tier with max-biased RSRP). Frequency band $F_1$ is used for both macro and micro UEs, while $F_2$ is dedicated only to biased UEs to avoid strong interference from the macro tier. Thus, the macro tier uses only $F_1$, while the micro tier uses both $F_1$ and $F_2$ for separate groups of UEs. Note that in this case, macro UEs and micro UEs suffer from both cross-tier and co-tier interference, while biased UEs only suffer from co-tier interference.
  \vspace{0.1cm}\item Max-SIR cell association: Max-SIR cell association scheme, where a UE always associates with the cell providing the highest SIR (it does not matter if it is a macro or a micro cell), has been studied in \cite{AnalysisMaxSIR},\cite{CoverageandRateInKTier},\cite{AnalysisMaxSINRConnectivity} with frequency reuse factor of 1. We adopt here max-SIR cell association with $K=1$ to represent the case of the scheme in \cite{AnalysisMaxSIR} where no frequency reuse is adopted and max-SIR with $K=2$ and $K=3$ to improve interference conditions.\\
\end{itemize}

For the sake of fair comparisons, we use the same numerical settings as in \cite{RateCoverate}, where $\lambda_M = 1$ $\V{M-eNBs/km}^2$, $P_{\mu} = 26$ dBm, and $\lambda_{\mu} = 5 \lambda_M$. The other parameters are the same as described in Section \ref{sec:Simulation}. With this configuration, according to \cite{RateCoverate}, the optimal bias value (RP2 case) of the micro tier to maximize the rate coverage probability is 15 dB and the fraction of bandwidth for biased UEs ($F_2$) is 0.47.

\begin{figure}
  \centering \epsfxsize=8cm \epsfbox{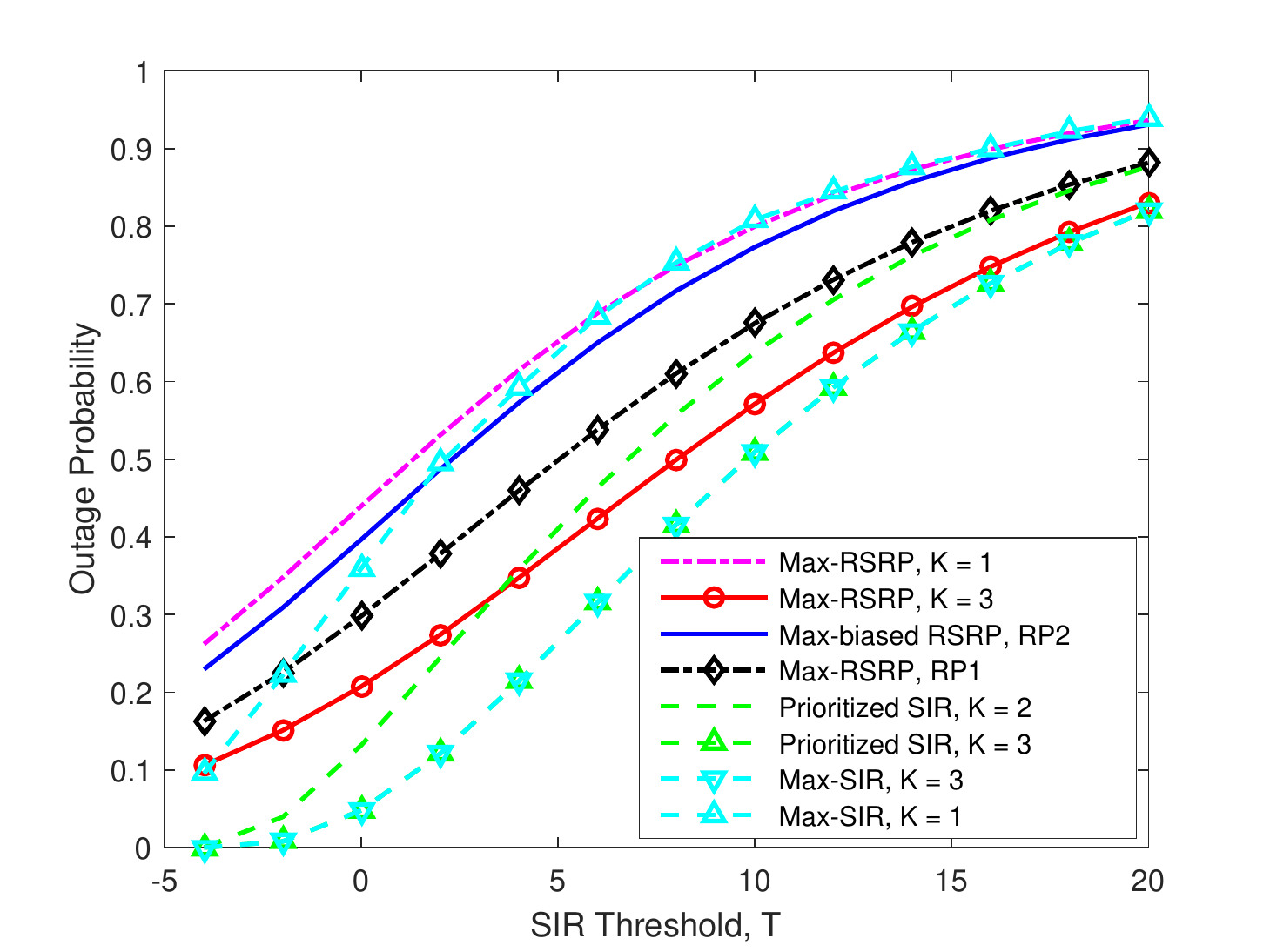}
    \caption{Outage probability comparisons.}\label{fig:OutageCompare}
\end{figure}

Figure \ref{fig:OutageCompare} shows the outage probability of our cell association scheme [denoted as ``prioritized SIR(-based)"] and of the other schemes in the literature obtained from simulations. We can see that our cell association scheme with $K = 3$ achieves much lower outage probability than the other RSRP-based schemes. This is because in our scheme UEs associate with cells providing SIR greater than the minimum SIR threshold $T$. Max-RSRP with $K=1$ (spectrum sharing) is characterized by the highest outage probability because of the high interference: all cells use the same frequency band. Max-biased RSRP cell association scheme with RP2 is also quite bad at managing interference, because both micro and macro UEs suffer from cross-tier and co-tier interference; using a larger bias value can also cause a higher outage probability because of co-tier interference experienced by biased UEs. With max-RSRP and RP1 scheme, macro and micro tiers use different frequency segments and UEs also associate with cells providing the best received power, thus having lower outage probability since cross-tier interference is eliminated. Finally, max-SIR scheme achieves the same outage probability as our scheme for $K = 3$, \myhl{because the prioritized SIR-based scheme does not change the coverage condition of max-SIR.} On the other hand, max-SIR has a much worse outage probability for $K = 1$; this is a further proof that the reuse of resources is a very important strategy for reducing outage probability.

\begin{figure}
  \centering \epsfxsize=7.5cm \epsfbox{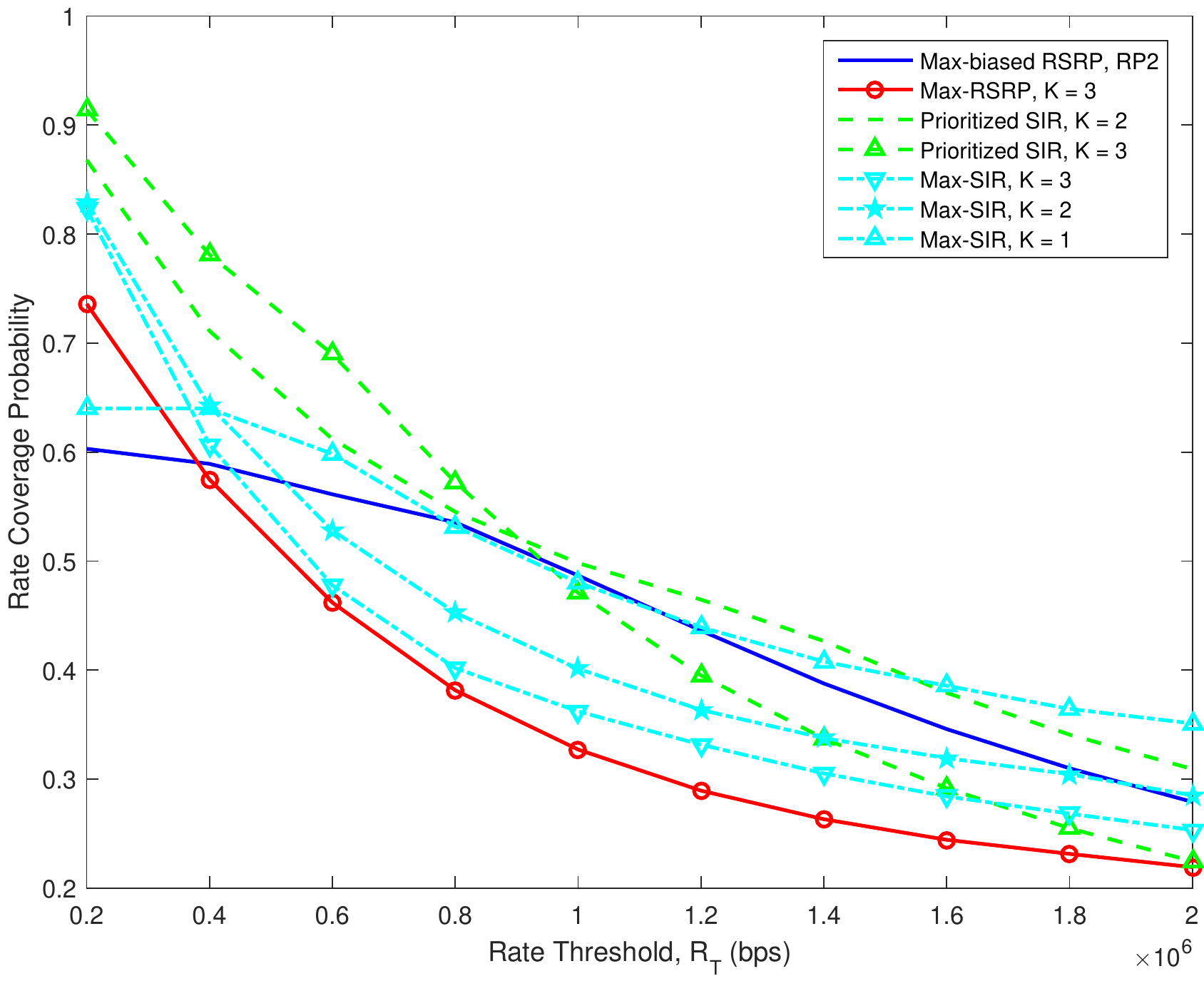}
  \caption{Rate coverage probability comparisons.}\label{fig:RateCompare}
\end{figure}

The rate coverage probabilities of the different schemes are shown in Fig. \ref{fig:RateCompare} (\footnote{\myhl{Note that the study in \cite{RateCoverate} for max-biased RSRP considers that UEs in outage can still have some throughput. This is different from our study where UEs in outage have no throughput, so that the rate coverage probability cannot be higher than the cell coverage probability $ {P}_c$.}}). We can see that our scheme outperforms max-RSRP with $K = 3$.
This occurs because max-RSRP causes most of the UEs to associate with the macro tier, so that macro UEs suffer from lack of resources (they cannot attain the required bit-rate). Instead, other schemes (such as max-biased RSRP and our prioritized SIR-based schemes) include load balancing mechanisms, so that they can achieve better performance. Our scheme with $K = 3$ outperforms max-biased RSRP RP2 scheme when $R_T < 1$ Mbps, because our scheme provides better SIR (see Fig. \ref{fig:OutageCompare}) and \myhl{exploits better the capacity of micro cells (higher $A_{\mu}$ value)}, thus UEs in edge areas can satisfy the rate requirement. For $R_T \geq 1$ Mbps, prioritized SIR-based scheme with $K = 3$ has a lower rate coverage probability because of the smaller amount of bandwidth available for each UE\footnote{With $K$ = 3, we have better SIR for UEs, thus the number of UEs in the coverage area is large. However, since each cell uses only 1/3 of the bandwidth (frequency reuse of 3), the bandwidth (bit-rate) for each UE of the cell is small. Thus, if $R_T$ is too large (and the threshold for this is about 1 Mbps), many UEs cannot satisfy the rate requirement so that rate coverage probability is low.}. Then, it would be better to use $K  = 2$ with our scheme for $R_T \geq 1$ Mbps in order to outperform max-biased RSRP. Moreover, our prioritized SIR-based scheme also obtains better rate coverage probability than max-SIR scheme for a wide range of $R_T$ values for both $K = 2$ and $K = 3$; in the $K = 1$ this is true only for small-to-medium $R_T$ values. These results of the comparison between our scheme and max-SIR depending on $K$ can be justified because more UEs associate with the micro tier (denoted below as `micro UEs') with prioritized SIR-based scheme than with max-SIR, thus spending the radio resources of the macro tier for those UEs that can only associate with macro cells. On the other hand, max-SIR is better than prioritized-SIR using the same $K$ value when $R_T$ is very high. \myhl{This occurs because micro UEs have higher bit-rates with max-SIR rather than with prioritized SIR-based scheme. In fact, prioritized-SIR causes more UEs to be associated with micro cells than with max-SIR. Thus, each micro UE can only have a smaller amount of bandwidth and its bit-rate is lower. Hence, only max-SIR can allow to satisfy the rate coverage probability requirement with high $R_T$ values.}

\subsection{Optimization Approach for Cell Planning}

Let us study how cellular network planning could be carried out  for our prioritized SIR-based cell association scheme. We propose an optimization approach based on the analysis that has been validated in a previous sub-Section.

We consider that the SIR threshold $T$ value is determined by the technology of the air interface, $\lambda_u/\lambda_M$ is given and is related to the cost of macro installations with respect to the expected revenue per user, $P_{\mu}/P_M$ is set depending on the air interface technology, $R_T$ is also given and pertains quality of experience. Then, the optimization problem has to determine $K$ and $\lambda_{\mu}/\lambda_M$ according to a criterion that reduces the costs related to the installations of $\mu$-eNBs under certain requirements on outage probability and rate coverage probability as follows:
\begin{equation}\label{def_optim}
\begin{array}{l}
 \mathop {\min }\limits_{K,\frac{{\lambda _\mu  }}{{\lambda _M }}} \frac{{\lambda _\mu  }}{{\lambda _M }} \\
 s.t. \\
 O < 10\%  \\
 P\left\{ {R > R_{T} } \right\} \ge 50\%,  \\
 \end{array}
\end{equation}
\noindent where $T$, $\lambda_u / \lambda_M$, $P_M/P_\mu$, $\gamma$, $W$ and $R_{T}$ are given.\\

In this study, $P\left\{ {R > R_{T} } \right\}$ is given by (\ref{eq:rateCoverFinal}). We assume that the above rate coverage constraint is feasible and defines a curve $\Gamma : P\left\{ {R > R_{T} } \right\} - 0.5 = 0$ in the $K$ - $\lambda_{\mu}/\lambda_M$ plane. This is verified under a certain range of $R_T$ values; if higher $R_T$ values have to be guaranteed, a larger $W$ and/or a lower $\lambda_u/\lambda_M$ are needed. If the planning has to guarantee a better performance it is possible to increase the $R_T$ value and/or to consider a larger value than 50\% for the rate coverage probability constraint.

The outage probability constraint $O < 10\%$ can be further elaborated according to (\ref{eq:coverageProbFinal}) as a function of $P_{c,1}(T)$ given in (\ref{eq:theorem1}). In particular, we have:
\begin{equation}
K > \frac{{ - 1}}{{\log _{10} \left( {1 - D\left( {\gamma ,T} \right)} \right)}}.
\end{equation}

Let us denote $d \triangleq \left\lfloor { - \left[ {\log _{10} \left( {1 - D\left( {\gamma ,T} \right)} \right)} \right]^{ - 1} } \right\rfloor$; then, the set of $K$ values fulfilling the above outage constraint is \{$d$+1, $d$+2, ...\}.

Other optimization approaches based on the maximization of the mean UE bit-rate could be considered, but the mean UE bit-rate increases with $\lambda_{\mu}/\lambda_M$ and then a limit due to cost constraints should be adopted for $\lambda_{\mu}/\lambda_M$ values and in most of the cases the optimization would just imply to select the maximum allowed $\lambda_{\mu}/\lambda_M$ value. This is the reason why these approaches are less interesting. As an alternative, we could substitute the rate coverage probability constraint with a constrain on the mean UE bit-rate; doing this way we would obtain a problem that can be solved with the same approach shown below, but with more complexity due to the integration needed to obtain the mean UE bit-rate. This is the reason why we do not deal with this case here.
\begin{figure*}[!htbp]
\begin{equation}\label{punto_selezione}
\left\{ \begin{array}{l}
 \frac{\partial }{{\partial K}}P\left( {R > R_T } \right) = 0 \\
 P\left( {R > R_T } \right) - 0.5 = 0 \\
 \end{array} \right.\quad  \Leftrightarrow \quad \left\{ \begin{array}{l}
 \frac{\partial }{{\partial K}}\left\{ {\left[ {1 - P_{c,1,\mu } \left( {T_\mu  } \right)} \right]^K  - \left[ {1 - P_{c,1,\mu } \left( T \right)} \right]^K  + \left[ {1 - P_{c,1} \left( {T_M ,T} \right)} \right]^K } \right\} = 0 \\
 0.5 - \left[ {1 - P_{c,1,\mu } \left( {T_\mu  } \right)} \right]^K  + \left[ {1 - P_{c,1,\mu } \left( T \right)} \right]^K  - \left[ {1 - P_{c,1} \left( {T_M ,T} \right)} \right]^K = 0 \\
 \end{array} \right.
\end{equation}
\end{figure*}

\begin{figure}
  \centering \epsfxsize=8cm \epsfbox{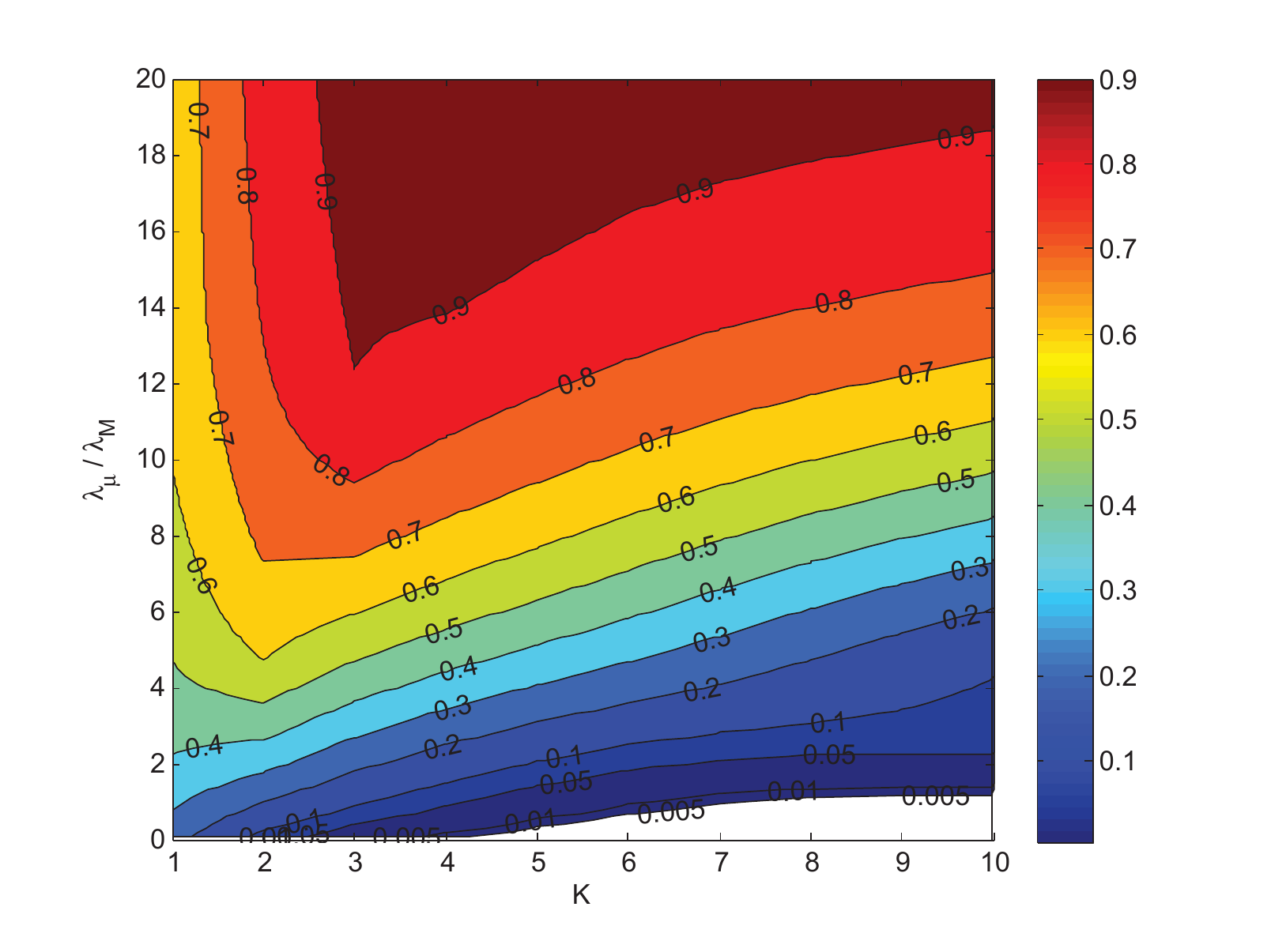}
  \caption{Contour plot of the rate coverage probability as function of $K$ and $\lambda_{\mu}/\lambda_M$.}
  \label{fig:ratecoverage}
\end{figure}

Since $K$ is integer, the optimization in (\ref{def_optim}) is a mixed-integer programming problem that is basically non-convex. The classical approach for solving this problem requires to find the optimal solution considering $K$ as a  continuous variable (relaxation) and then applying the Branch and Bound method. This optimization has been solved by means of a heuristic approach that can be explained using the graphical example in Fig. \ref{fig:ratecoverage}, showing level curves of the rate coverage probability in the $K$ - $\lambda_{\mu}/\lambda_M$ plane for $R_T$ = 1 Mbit/s and $P_\mu$ = 30 dBm. In this figure, we have to select the point in the plane with the lowest $\lambda_{\mu}/\lambda_M$ value that is fulfilling both the outage constraint (i.e., $K \ge 3$, $d$ = 2) and the rate coverage probability constraint. Because of the behavior of the rate coverage probability, this point can be found on the level curve $\Gamma$ (i.e., rate coverage probability equal to 0.5) as the point with the lowest $\lambda_{\mu}/\lambda_M$ value for integer $K \ge 3$. Then, in our example, the optimum point is achieved for $K$ = 3 and $\lambda_{\mu}/\lambda_M \approx 5$. This graphical solution suggests the following method to solve our problem.

We consider that the relaxed constraints (corresponding to curves $\Gamma$ and $K > d$) define a \emph{feasibility area} in the $K$ - $\lambda_{\mu}/\lambda_M$ plane where we have to find the point with minimum $\lambda_{\mu}/\lambda_M$. Then, as shown in the example in Fig. \ref{fig:ratecoverage}, the solution point is on the border of the area on the $\Gamma$ curve; this is an implicit function of $K$ for which we can find the minimum by means of the null derivative condition in (\ref{punto_selezione}) at the top of this page \cite{Dini}, where unknown variables are $K$ and $\lambda_{\mu}/\lambda_M$ and the other parameters are given. Because of the complexity in formally expressing the solution of (\ref{punto_selezione}), this system has been solved numerically. We have two cases:\\

\begin{itemize}[leftmargin=.2in]
\item If (\ref{punto_selezione}) has a solution $K^*$, then we consider the corresponding closer integer values of $K$, denoted as $\left\lfloor {K^* } \right\rfloor$ and $\left\lfloor {K^*} \right\rfloor +1$, and we choose among them as follows:
$$
\left\{ \begin{array}{l}
{\rm if}\quad d + 1 \le \left\lfloor {K^* } \right\rfloor \quad  \Rightarrow {\rm  we ~select}~K = \left\lfloor {K^* } \right\rfloor {\rm   } \\
 \quad \;\;{\rm or}~K = \left\lfloor {K^* } \right\rfloor  + 1~ {\rm  depending~ on ~whichever ~of ~the } \\
 \quad \;\;{\rm two ~allows ~the ~lower} ~\frac{{\lambda _\mu  }}{{\lambda _M }}{\rm ~value ~on} ~\Gamma;  \\
{\rm if}\quad \left\lfloor {K^* } \right\rfloor  + 1 \le d + 1\quad  \Rightarrow {\rm  we ~select} ~K = d + 1{\rm  ~and } \\
 \quad \;\;{\rm the ~corresponding} ~\frac{{\lambda _\mu  }}{{\lambda _M }}{\rm  ~value ~on} ~\Gamma.  \\
 \end{array} \right.
$$

\item If (\ref{punto_selezione}) has no solution, we just select $K$ = $d$ + 1 and the corresponding $\lambda_{\mu}/\lambda_M$ value on the $\Gamma$ curve.\\
\end{itemize}

\begin{figure}
  \centering \epsfxsize=8cm \epsfbox{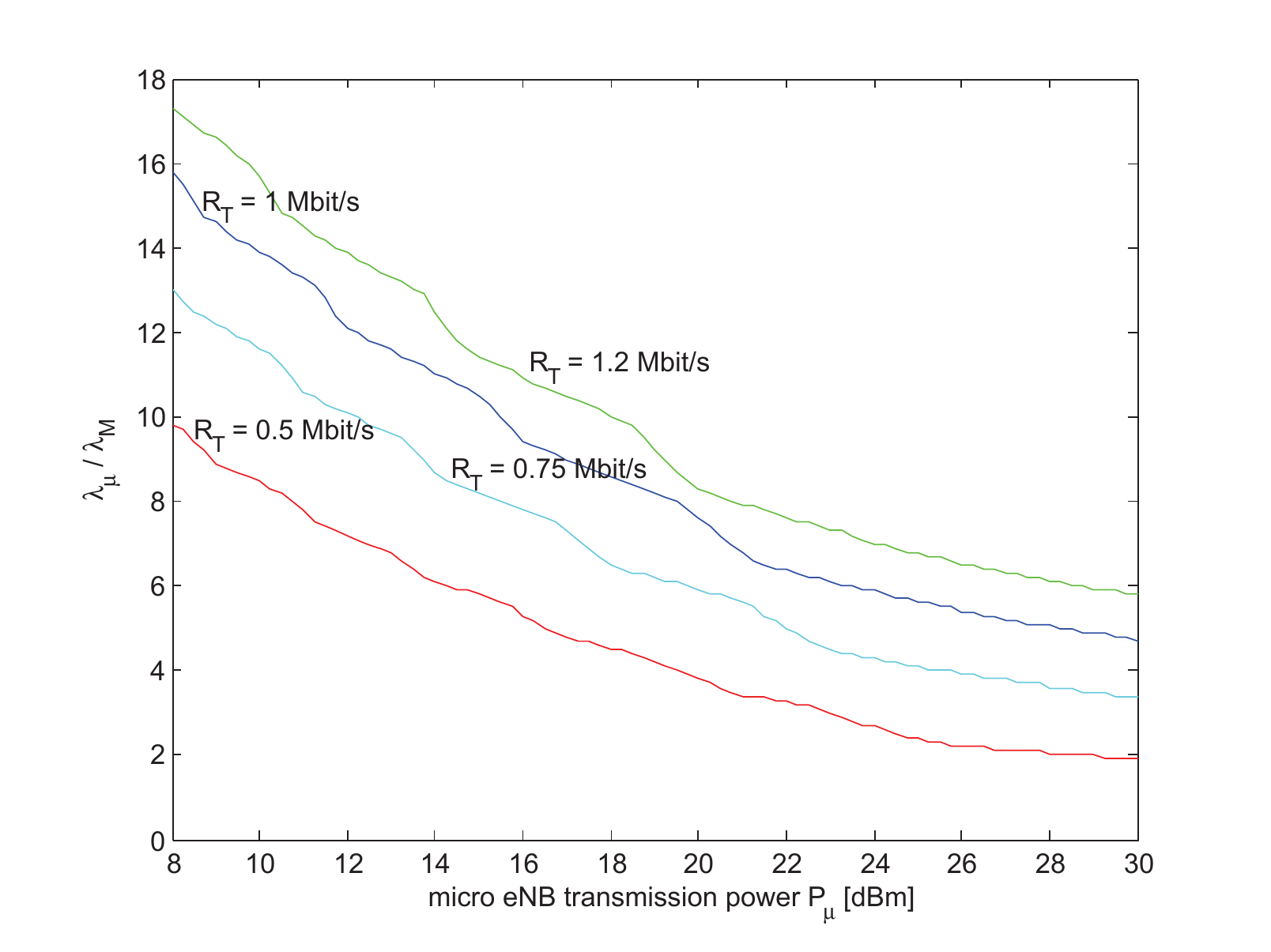}
  \caption{Optimized $\lambda_{\mu}/\lambda_M$ as a function of $P_\mu$ for different $R_T$ values.}\label{fig:selectedlamdda}
\end{figure}

\begin{figure}
  \centering \epsfxsize=8cm \epsfbox{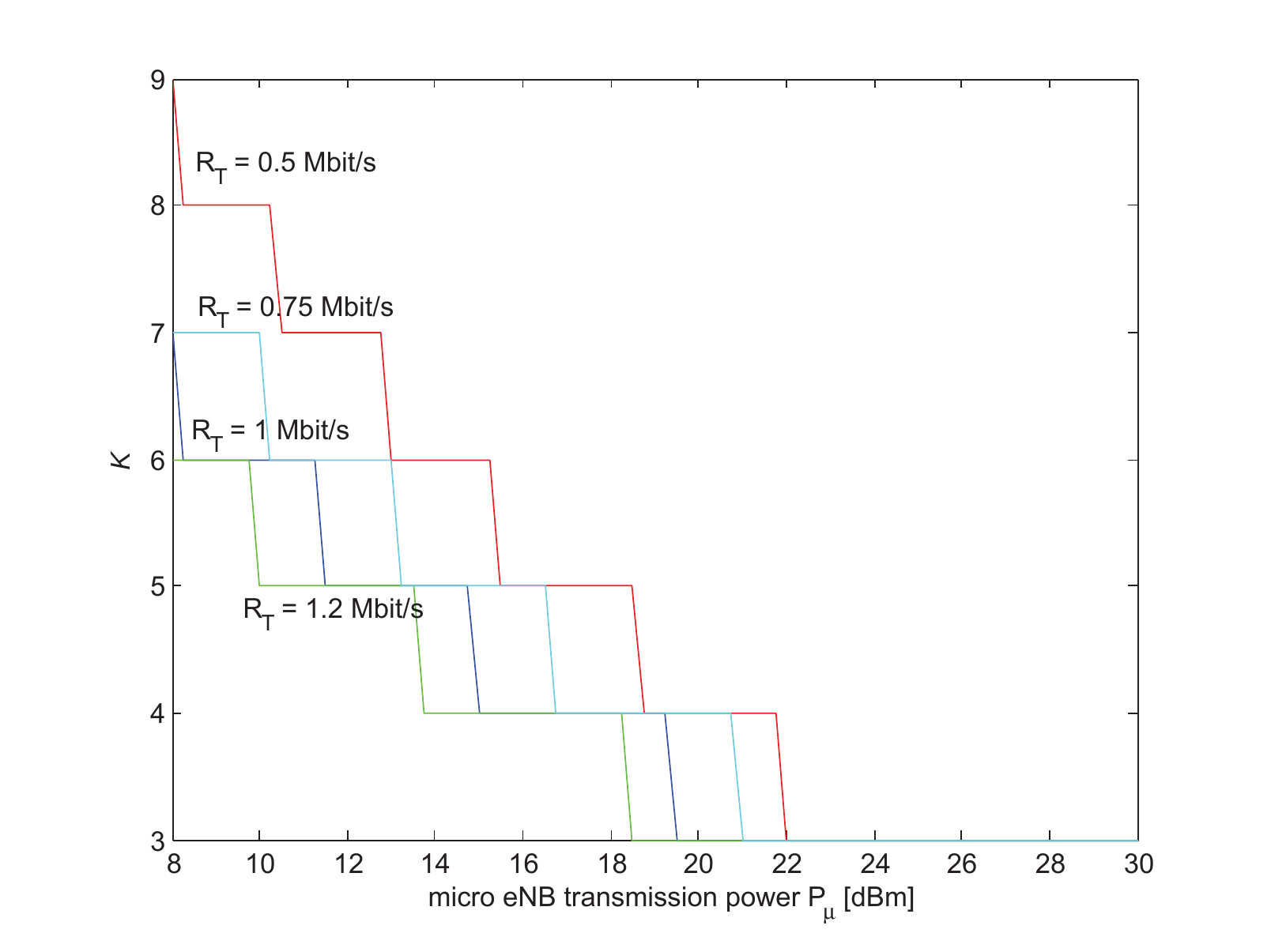}
  \caption{Optimized $K$ as a function of $P_\mu$ for different $R_T$ values.}\label{fig:selectedK}
\end{figure}

By applying our optimization approach, Figs. \ref{fig:selectedlamdda} and \ref{fig:selectedK} show the selected $K$ and $\lambda_{\mu}/\lambda_M$ values for different $P_\mu$ values (and  $P_M$ = 46 dBm). We can see that the optimized values of $\lambda_{\mu}/\lambda_M$ reduce with $P_\mu$, because increasing the transmission power of $\mu$-eNB we enlarge the micro cell service area; a similar effect on the optimized values of $\lambda_{\mu}/\lambda_M$ can be obtained by reducing $R_T$. Moreover, the selected $K$ values reduce with $P_\mu$, because the rate coverage constraint is satisfied with lower $K$ values if $P_\mu$ increases; starting from a certain $P_\mu$ value, we can use the lowest (integer) $K$ value fulfilling the outage probability constraint, that is $d+1$ with our notations. By means of (\ref{eq:rateCoverFinal2}) we can also compute the mean UE bit-rate corresponding to the different optimized configurations depending on $\mu$-eNB transmission power: we have that the mean UE bit-rate decreases with $P_\mu$ (or, equivalently, increases with $\lambda_{\mu}/\lambda_M$) and increases with $R_T$. Hence, a good planning approach based on Figs. \ref{fig:selectedlamdda} and \ref{fig:selectedK} should select a lower $P_\mu$ value as much as possible considering the cost of a higher $\mu$-eNB density.



\vspace{10pt}
\section{\uppercase{Conclusions}}
\label{sec:Conclusions}
Using small cells in combination with macro ones is a key approach towards 5G cellular systems. In this paper, we have presented an analytical framework to characterize the performance of HetNets with stochastic geometry, frequency reuse of $K$, and prioritized SIR-based cell association scheme. Analytical results have been achieved that allow us to characterize outage probability, average load, rate coverage probability, and mean UE bit-rate. Simulation and analytical results show that our prioritized SIR-based cell association scheme can obtain low outage probability values, while providing better rate coverage probability than other key schemes in the literature. The study carried out in this paper provides a closed-form analytical method that may help network operators when planning future cellular systems.

A further work will be needed to provide an analytical framework for other frequency reuse schemes, such as Soft Frequency Reuse. Finally, more than two tiers (e.g., femtocells) will be included in a future study.

\vspace{10pt}
\section*{\uppercase{Appendix A}}
\label{appendixA}

\subsection*{Proof of Theorem \ref{theremOutage}}

We provide the proof of Theorem \ref{theremOutage}. Out of all the cells of the cellular system, we consider here only those cells of the different tiers where $F_1$ is used (considering any other frequency segment would be the same). Let $\Phi_{M,1}$, $\Phi_{\mu,1}$ be the sub-sets of M-eNBs and \textmu-eNBs, which use frequency $F_1$. $ {P}_{c,1}(T_M, T_{\mu})$ denotes the coverage probability for the reference UE in the origin and only referring to those cells using $F_1$ and adopting threshold $T_M$ for macro and threshold $T_{\mu}$ for micro cells.

The following derivations are carried out in the most general case with background noise with power $\sigma^2$ so that we refer here to SINR rather than to SIR. Note that the derivations in this Appendix are obtained by modifying the approach shown in \cite{AnalysisMaxSIR} to take the reuse of frequency with factor $K$ into account. In particular, we can express $ {P}_{c,1}(T_M, T_{\mu})$ as follows:

\begin{equation}\label{th:coverageProb}
  \begin{split}
      {P}_{c,1}(T_M, T_{\mu}) & =  {P} \left( \bigcup_{i \in \Phi_{M,1} \cup \Phi_{\mu,1}} \{\V{SINR}_i  \geq T_i \} \right) \\
       & =  {P} \left( \bigcup_{i \in \Phi_{M,1} \cup \Phi_{\mu,1}} B_i\right), 
  \end{split}
\end{equation}
where $B_{i}$ is the event $\{\V{SINR}_i \geq T_i\}$ and where $T_i$ has one of the two values $T_M$ or $T_{\mu}$, depending on subscript $i$ corresponding to a M-eNB or to a \textmu-eNB.
We apply Lemma 1 of \cite{AnalysisMaxSIR} to solve (\ref{th:coverageProb}). In particular, when $T_i \geq 1$ (0 dB) and $K$ = 1, Lemma 1 states that $ {P}(B_i\cap B_j) = 0$, i.e., events $B_i$ and $B_j$ cannot happen at the same time (the coverage areas of different cells are disjoint). We can explain this fact as follows. Let us consider a simple scenario where there are three signals from three eNBs using frequency $F_1$ that are received at the reference UE with power levels $b_1$, $b_2$, and $b_3$ ($b_1, b_2, b_3 > 0$). We assume that SIR threshold $T$ is 1 (0 dB). Then, if $\V{SIR}_1$ = $\frac{b_1}{b_2 + b_3} > 1$ from eNB \#1, we have $b_1 > b_2 + b_3$, which leads to $\V{SIR}_2 = \frac{b_2}{b_1 + b_3} < 1$ from eNB \#2 and $\V{SIR}_3 = \frac{b_3}{b_1 + b_2} < 1$ from eNB \#3. Thus, the reference UE can experience a SIR greater than $T = 1$ from at most one eNB, that is eNB \#1 in this example. This result can be extended to the cases $T \geq 1$. Using the same reasoning, we can also prove that the coverage areas are not disjoint when $T < 1$ (0 dB). In order to solve (\ref{th:coverageProb}) and to simplify the analysis, we consider that cell coverage areas are always disjoint, even if this is true only as a first approximation when $T < 1$ (0 dB). Note that, power $K$ in formula (\ref{eq:coverageProbFinal}) reduces the effects of such approximation so that the differences between analysis and simulations are very small when $K > 1$ (see Fig. \ref{fig:OutagevsSIR}). With $B_i$'s being disjoint events, the probability of the union of events in (\ref{th:coverageProb}) is equal to the sum of individual probabilities as follows:

\begin{equation}\label{th:coverageProbReduced}
  \begin{split}
      {P}_{c,1}(T_M, T_{\mu}) & =  {P} \left( \bigcup_{i \in \Phi_{M,1} \cup \Phi_{\mu,1}} B_i\right) \approx  {E} \left[ \sum_{i \in \Phi_{M,1} \cup \Phi_{\mu,1}}  {P}(B_i) \right] \\
        & = \sum_{j = \{M,\mu\}}  {E} \left[\sum_{i \in \Phi_{j,1}}  {P}(B_i) \right].
  \end{split}
\end{equation}
where operator $E[.]$ is over $\Phi$ processes.


\myhl{Note that this approximation provides a upper bound to the coverage probability.} Let us consider a stationary point process $\Phi$ with constant density $\lambda$ and a function $f: {\mathbb{R}}^2 \rightarrow {\mathbb{R}}$ that is applied to the generic point $x \in \Phi$ in $ {\mathbb{R}}^2$. Then, applying the Campbell-Mecke Theorem \cite{StochasticGeoandApplications}, we have:
\begin{equation}\label{eq:campbellMecke}
  \mathbf{E} \left[ \sum_{x \in \Phi} f(x)\right] = \lambda \int_{ {\mathbb{R}}^2} f(x) \V{d}x.
\end{equation}
 Then, in our case, $x$ is the generic position of macro or micro eNBs that use $F_1$ according to sub-process $\Phi_{j,1}$. Moreover, $\lambda$ is the density of macro or micro cells of the sub-process $\Phi_{j,1}$, being $\lambda$ equal to $\frac{\lambda_j}{K}$ since the process with density $\lambda_j$ is split in $K$ parts because of the division of the spectrum in $K$ parts \cite{StochasticSurvey},\cite{StochasticGeometryTheory}. $f(x)$ corresponds to the probability $ {P}(B_j) =  {P}\left(\V{SINR}_j \geq T_j\right)$. Thus, we obtain:
\begin{equation}\label{th:campbell}
\begin{split}
  &  {P}_{c,1}(T_M, T_{\mu}) \approx \sum_{\underset{}{j = \{M, \mu\}}} \frac{\lambda_j}{K} \int_{ {\mathbb{R}}^2}  {P} \left( \V{SINR}_j \geq T_j \right) \V{d}r_j \\
     &= \sum_{\underset{}{j = \{M, \mu\}}} \frac{\lambda_j}{K} \int_{ {\mathbb{R}}^2}  {P} \left( \frac{P_j h_j \parallel r_j\parallel^{-\gamma}}{I_j + \sigma^2} \geq T_j \right) \V{d}r_j,
\end{split}
\end{equation}
where $\parallel r_j \parallel$ denote the distances between the reference UE and eNBs in tier $j$ (points $r_j$ correspond to points $x$ according to the previous notation), $I_j$ denotes the interference power received from other eNBs in tier $j$ using $F_1$, and $\sigma^2$ denotes the noise power.

By using the same approach as that in \cite{AnalysisMaxSIR}, we obtain the expression of $ {P}_{c,1}(T_M, T_{\mu})$ with frequency reuse of $K$ as shown in (\ref{th:generalizedCoverageProb}) at the top of the next page, where $C(\gamma)= \frac{2 \pi^2}{\gamma \sin(\frac{\pi}{\gamma/2})}$. If noise is ignored ($\sigma^2 = 0$), and changing to polar coordinates, we have: $\parallel r_j \parallel^2 = \upsilon^2$, $\V{d}r_j = \upsilon \V{d}\upsilon \V{d}\theta$. Then, equation (\ref{th:generalizedCoverageProb}) becomes equation (\ref{th:noiseIgnored}), as shown at the top of the next page;
\begin{figure*}
\begin{equation}\label{th:generalizedCoverageProb}
\begin{split}
    {P}_{c,1}(T_M, T_{\mu})
    = \sum_{\underset{}{j \approx \{M, \mu\}}} \frac{\lambda_j}{K} \int_{ {\mathbb{R}}^2} \exp\left[ -\left(\frac{T_j}{P_j}\right)^{2/\gamma} \parallel r_j \parallel^2 C(\gamma) \sum_{m = \{M,\mu\}} \frac{\lambda_m}{K} P_m^{2/\gamma}\right]
      \exp\left(\frac{-T_j \sigma^2 \parallel r_j \parallel^\gamma}{P_j}\right) \V{d}r_j.
\end{split}
\end{equation}
\end{figure*}
\begin{figure*}
\begin{equation}\label{th:noiseIgnored}
\begin{split}
        {P}_{c,1}(T_M, T_{\mu}) & \approx  \sum_{j = \{M, \mu\}} \frac{\lambda_j}{K} \int_{0}^{+\infty}\int_{0}^{2\pi} \exp\left[ -\left(\frac{T_j}{P_j}\right)^{2/\gamma} \upsilon^2 C(\gamma) \sum_{m = \{M,\mu\}} \frac{\lambda_m}{K} P_m^{2/\gamma}\right] \upsilon \V{d}\upsilon \V{d}\theta  \\
   &= \sum_{j = \{M, \mu\}} \frac{\lambda_j}{K} \int_{0}^{2\pi} \V{d}\theta \int_{0}^{+\infty} \exp\left[ -\left(\frac{T_j}{P_j}\right)^{2/\gamma} \upsilon^2 C(\gamma) \sum_{m = \{M,\mu\}} \frac{\lambda_m}{K} P_m^{2/\gamma}\right] \frac{1}{2}\V{d}\upsilon^2 \\
   &=  \frac{\pi \sum_{j = \{M, \mu\}} \lambda_j P_j^{2/\gamma} T_j^{-2/\gamma}}{C(\gamma) \sum_{m = \{M,\mu\}} \lambda_m P_m^{2/\gamma}} =  D(\gamma, T)\frac{{\left( {\frac{{T_\mu  }}{T}} \right)^{ - \frac{2}{\gamma }}  + \left( {\frac{{\lambda _M }}{{\lambda _\mu  }}} \right)\left( {\frac{{P_M }}{{P_\mu  }}} \right)^{\frac{2}{\gamma }} \left( {\frac{{T_M }}{T}} \right)^{ - \frac{2}{\gamma }} }}{{1 + \left( {\frac{{\lambda _M }}{{\lambda _\mu  }}} \right)\left( {\frac{{P_M }}{{P_\mu  }}} \right)^{\frac{2}{\gamma }} }}.
\end{split}
\end{equation}
\end{figure*}
if $T_M = T_{\mu} = T$, we obtain:
\begin{equation}\label{th:finalProof}
    {P}_{c,1} =  {P}_{c,1}(T, T) \approx D(\gamma, T) \triangleq \frac{\pi}{C(\gamma)T^{2/\gamma}},
\end{equation}
which completes our proof.

\vspace{10pt}
\section*{\uppercase{Appendix B}}
\label{appendixB}
\subsection*{Proof of Proposition 1}

Let $ {P}_{c,1,\mu} (T_{\mu})$ denote the coverage probability (for the reference UE in the origin) of the micro tier with SINR threshold $T_{\mu}$ when only the subset of micro cells using frequency $F_1$ is considered. We have:

\begin{equation}\label{pr:microCoverage}
   {P}_{c,1,\mu} (T_{\mu}) =   {P} \left( \bigcup_{i \in  \Phi_{\mu,1}} \{\V{SINR}_i  \geq T_{\mu} \} \right).
\end{equation}

By using the same method as in Appendix A and assuming no noise, we obtain the formula of $ {P}_{c,1,\mu} (T_{\mu})$ as follows:

\begin{equation}\label{pr:micCoverage1Fre}
\begin{split}
   {P}_{c,1,\mu} (T_{\mu}) &= \frac{\lambda_{\mu}\pi P_{\mu}^{2/\gamma} T_{\mu}^{-2/\gamma}}{C(\gamma) \sum_{m = \{M,\mu\}} \lambda_m P_m^{2/\gamma} }\\
   & = D(\gamma,T)\frac{{\left( {\frac{{T_\mu  }}{T}} \right)^{- \frac{2}{\gamma }} }}{{1 + \left( {\frac{{\lambda _M }}{{\lambda _\mu  }}} \right)\left( {\frac{{P_M }}{{P_\mu  }}} \right)^{\frac{2}{\gamma }} }}.
\end{split}
\end{equation}
If we consider $T_{\mu} = T$, we obtain the expression of $ {P}_{c,1,\mu}$ that is used in Proposition \ref{proprosionAverageLoad}. Moreover, by means of the same approach as in sub-Section \ref{subsec:OutageProb}, we obtain the coverage probability $ {P}_{c,\mu}$ of the micro tier as:

\begin{equation}\label{pr:finalMicCoverage}
    {P}_{c,\mu} = 1 - \left( 1 -  {P}_{c,1,\mu} \right)^K.
\end{equation}

Now let us derive the average load on the micro tier. According to the prioritized SIR-based cell association scheme, the reference UE will associate with the micro tier if it experiences a SIR value higher than threshold $T$ from at least one cell of this tier. Thus, we have:
\begin{equation}\label{pr:microcellAssoc}
\begin{split}
   A_{\mu} & =  {P} \left( \bigcup_{j \in  \Phi_{\mu} } \{\V{SIR}_j  \geq T\} | \bigcup_{i \in  \Phi_{M} \cup \Phi_{\mu}} \{\V{SIR}_i  \geq T\} \right) \\
     & = \frac{ {P} \left( \bigcup_{j \in  \Phi_{\mu} } \{\V{SIR}_j  \geq T \}, \bigcup_{i \in  \Phi_{M}\cup \Phi_{\mu}} \{\V{SIR}_i  \geq T \}\right) }{ {P} \left(\bigcup_{i \in  \Phi_{M} \cup \Phi_{\mu}} \{\V{SIR}_i  \geq T \}\right)} \\
     & = \frac{ {P} \left( \bigcup_{j \in  \Phi_{\mu} } \{\V{SIR}_j  \geq T\} \right) }{ {P} \left(\bigcup_{i \in  \Phi_{M}\cup \Phi_{\mu}} \{\V{SIR}_i  \geq T \}\right)} \\
     & = \frac{ {P}_{c,\mu}}{ {P}_{c}} = \frac{1-(1- {P}_{c,1,\mu})^K}{1-(1- {P}_{c,1})^K},
\end{split}
\end{equation}
where at the third step we have simplified the joint probability because $ \bigcup_{j \in  \Phi_{\mu} } \{\V{SIR}_j  \geq T\}$ is a subset of $\bigcup_{i \in  \Phi_{M}\cup \Phi_{\mu}} \{\V{SIR}_i  \geq T \}$.\\

\noindent This completes the proof.


\vspace{10pt}
\bibliographystyle{IEEE}

\epsfysize=3.2cm
\begin{biography}{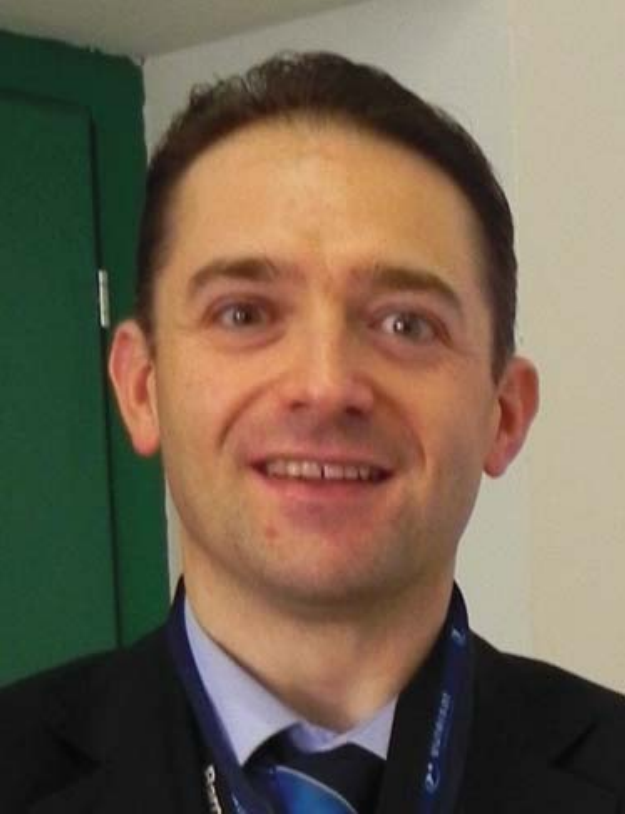}{Giovanni Giambene} received the Dr. Ing. degree in Electronics in 1993 and the Ph.D. degree in Telecommunications and Informatics in 1997, both from the University of Florence, Italy. He was Technical External Secretary of the European Community COST 227 Action (“Integrated Space/Terrestrial Mobile Networks''). From 1997 to 1998, he was with OTE (Marconi Group) in Florence, Italy, working on a GSM development program. In 1999, he joined the Department of Information Engineering and Mathematical Sciences of the University of Siena, Italy, where, he is associate professor on Networking. He was vice-Chair of the COST 290 Action (2004-2008), entitled “Traffic and QoS Management in Wireless Multimedia Networks'' (Wi-QoST). He participated to: (i) the SatNEx I \& II network of excellence (EU FP6, 2004-2009) and SatNEx III (ESA 2010-2013); (ii) the EU FP7 Coordination Action “Road mapping technology for enhancing security to protect medical \& genetic data'' (RADICAL); (iii) the COST Action IC0906 (2010-2014) “Wireless Networking for Moving Objects'' (WiNeMO) as national representative; (iv) the EU FP7 Coordination Action RESPONSIBILITY coordination action (2013-2016). At present, he is involved in the ESA SatNEX IV project. Giambene is IEEE senior member. Since January 2015, he is IEEE Transactions on Vehicular Technology editor. His research interests deal with wireless and satellite networking, cross-layer air interface design, transport layer performance, privacy and security for medical data.
\end{biography}
\epsfysize=3.2cm
\begin{biography}{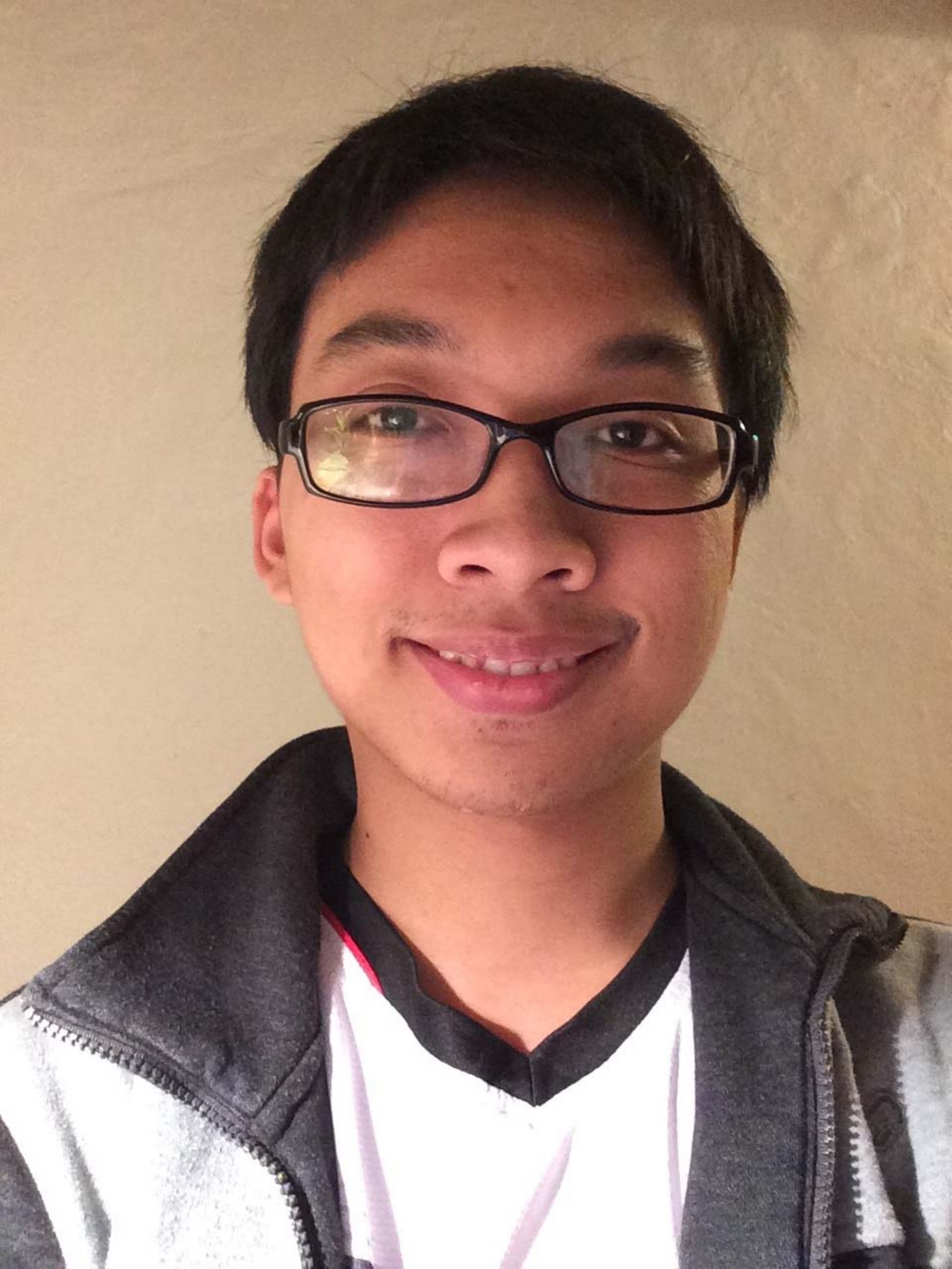}{Van Anh Le}
received the degree in Electronics and Telecommunications from the Hanoi University of Science and Technology, Vietnam in 2012 and the Ph.D. degree in Telecommunications from the University of Siena, Italy in 2016. His main research interests include resource management, cellular networks, and quality of service support.
\end{biography}

\end{document}